\global\def\draftcontrol{0}

   \def\versionno{ kt-hydro }

\catcode`\@=11

\expandafter\ifx\csname draftcontrol\endcsname\relax\global\def\draftcontrol{0}
\fi

{\count255=\time\divide\count255 by 60
\xdef\hourmin{\number\count255}
\multiply\count255 by-60\advance\count255 by\time
\xdef\hourmin{\hourmin:\ifnum\count255<10 0\fi\the\count255}}
\def\draftdate{\number\month/\number\day/\number\year\ \ \ \hourmin }

\newcommand\makepapertitle{\par
  \begingroup
    \renewcommand\thefootnote{\@fnsymbol\c@footnote}%
    \def\@makefnmark{\rlap{\@textsuperscript{\normalfont\@thefnmark}}}%
    \long\def\@makefntext##1{\parindent 1em\noindent
            \hb@xt@1.8em{%
                \hss\@textsuperscript{\normalfont\@thefnmark}}##1}%
     \newpage
     \global\@topnum\z@   
     \@makepapertitle
     \thispagestyle{empty}\@thanks
  \endgroup
  \setcounter{footnote}{0}%
  \global\let\thanks\relax
  \global\let\makepapertitle\relax
  \global\let\@makepapertitle\relax
  \global\let\@thanks\@empty
  \global\let\@author\@empty
  \global\let\@date\@empty
  \global\let\@title\@empty
  \global\let\title\relax
  \global\let\author\relax
  \global\let\date\relax
  \global\let\and\relax
  \def\version{\let\version\@version\@gobble}
}
\def\@makepapertitle{%
  \newpage
   \ifnum\draftcontrol=1 {}
   \version\versionno
   \vskip 3em%
   \else
   \hfill\hbox to 3cm {\parbox{4cm}{\@pubnum}\hss}%
   \vskip 3em%
   \fi
   \begin{center}%
   \let \footnote \thanks
     {\LARGE {\@title}}%
     \vskip 1.5em%
     {\normalsize
       \lineskip .5em%
       \begin{tabular}[t]{c}%
         \@author
       \end{tabular}\par}%
     \vskip 1.5em%
     {\@bstract}%
     \end{center}%
     \vskip 1.5em
     \@date%
   \par
}

\gdef\@pubnum{}
\def\pubnum#1{%
  \gdef\@pubnum{#1}}

\gdef\@bstract{}
\def\Abstract#1{%
  \gdef\@bstract{%
   \parbox{\textwidth-0pc}{%
   \centerline{\bf Abstract}\penalty1000%
\kern.2cm%
\noindent
\renewcommand\baselinestretch{1.0}%
{#1}}}
}

\def\ps@paper{\let\@mkboth\@gobbletwo%
     \ifnum\draftcontrol=1
    \def\@oddfoot{\hbox to \textwidth{\tiny \versionno \hfil\tiny\draftdate}%
    \hskip -\textwidth \hbox to \textwidth{\hfil\rm\thepage\hfil}}%
     \else\def\@oddfoot{\hbox to \textwidth{\hfil\rm\thepage\hfil}}
     \fi
     \let\@evenfoot\@oddfoot
}

\def\body{\clearpage
          \pagestyle{paper}
    }

\def\@version#1{\ifnum\draftcontrol=1
\typeout{}\typeout{#1}\typeout{}
\vskip3mm\centerline{\hbox{\fbox{\normalsize{\tt DRAFT -- #1 -- }
                   {\draftdate}}}}\vskip3mm
\fi}
\let\version\@version
\long\def\eqlabel#1{\ifnum\draftcontrol=1
                    \tag@false  
                    \tag*{(\theequation) \hbox to -0.2cm{\hspace{0cm}\small{#1}\hss}}
                    \refstepcounter{equation}
                    \edef\@currentlabel{\theequation}
                    \ltx@label{#1}          
                    \else
                    \label{#1}
                    \fi
                    }
\let\st@bibitem\@bibitem
\let\st@lbibitem\@lbibitem
\ifnum\draftcontrol=1
  \def\@bibitem#1{%
    \st@bibitem{#1}\a@@label{#1}\ignorespaces}
  \def\@lbibitem[#1]#2{%
    \st@lbibitem[#1]{#2}\a@@label{#2}\ignorespaces}
  \def\a@@label#1{%
    \gdef\a@lab{\smash{\normalfont\small#1}}
    \ifvmode
      \if@inlabel
        \global\setbox\@labels\hbox{%
          \llap{\a@lab\let\a@lab\relax
                \kern\@totalleftmargin\kern\marginparsep}%
          \box\@labels}%
      \fi
    \fi}
\fi

\documentclass[12pt,letterpaper]{article}

\usepackage{psfrag,amsmath,amssymb,array,calc,epsfig,rotating,bm}
\usepackage[sort]{cite}

\ifnum\draftcontrol=1
\fi

\renewcommand\baselinestretch{1.25}
\renewcommand\section{\@startsection {section}{1}{\z@}%
                                   {-3.5ex \@plus -1ex \@minus -.2ex}%
                                   {2.3ex \@plus.2ex}%
                                   {\normalfont\large\bfseries}}
\renewcommand\subsection{\@startsection{subsection}{2}{\z@}%
                                   {-3.25ex\@plus -1ex \@minus -.2ex}%
                                   {1.5ex \@plus .2ex}%
                                   {\normalfont\normalsize\bfseries}}
\renewcommand\subsubsection{\@startsection{subsubsection}{3}{\z@}%
                                   {-3.25ex\@plus -1ex \@minus -.2ex}%
                                   {1.5ex \@plus .2ex}%
                                   {\normalfont\normalsize\it}}
\renewcommand\paragraph{\@startsection{paragraph}{4}{\z@}%
                                   {-3.25ex\@plus -1ex \@minus -.2ex}%
                                   {1.5ex \@plus .2ex}%
                                   {\normalfont\normalsize\bf}}

\numberwithin{equation}{section}

\def\revise#1       {\raisebox{-0em}{\rule{3pt}{1em}}%
                     \marginpar{\raisebox{.5em}{\vrule width3pt\
                     \vrule width0pt height 0pt depth0.5em
                     \hbox to 0cm{\hspace{0cm}{%
                     \parbox[t]{4em}{\raggedright\footnotesize{#1}}}\hss}}}}

\newcommand\nxt[1]  {\\\fnxt#1}

\def\cala         {{\cal A}}

\def\cale         {{\cal E}}
\def\calf         {{\cal F}}

\def\calj         {{\cal J}}

\def\caln         {{\cal N}}
\def\calo         {{\cal O}}
\def\calp         {{\cal P}}
\def\calq         {{\cal Q}}

\def\del          {\partial}

\def\tr           {\mathop{\rm Tr}}

\def\sqr#1#2{{\vcenter{\vbox{\hrule height.#2pt
 \hbox{\vrule width.#2pt height#1pt \kern#1pt
 \vrule width.#2pt}\hrule height.#2pt}}}}

\def\a{\alpha}
\def\b{\beta}
\newcommand{\qq}{\mathfrak{q}}
\newcommand{\ww}{\mathfrak{w}}

\newcommand{\beq}{\begin{equation}}
\newcommand{\eeq}{\end{equation}}
\newcommand{\beqa}{\begin{eqnarray}}
\newcommand{\eeqa}{\end{eqnarray}}
\newcommand{\beqar}{\begin{eqnarray*}}
\newcommand{\eeqar}{\end{eqnarray*}}

\renewcommand{\eqref}[1]{(\ref{#1})}

\newcommand{\ie}{{\it i.e.,}\ }

\def\a{\alpha}
\def\w{\omega}
\def\g{\gamma}

\def\dd{{\delta}}
\def\e{\epsilon}
\def\ta{\tilde{a}}
\def\k{\kappa}

\def\l{\lambda}
\def\z{\zeta}
\def\ha{\hat{a}}
\catcode`\@=12

\begin{document}


\title{\bf Hydrodynamics of the cascading plasma}
\pubnum{UWO-TH-09/6}


\author{
Alex Buchel\\[0.4cm]
\it Perimeter Institute for Theoretical Physics\\
\it Waterloo, Ontario N2L 2Y5, Canada\\[.5em]
 \it Department of Applied Mathematics\\
 \it University of Western Ontario\\
\it London, Ontario N6A 5B7, Canada
 }

\Abstract{
The cascading gauge theory of Klebanov {\it et.al} realizes a soluble
example of gauge/string correspondence in a non-conformal
setting. Such a gauge theory has a strong coupling scale $\Lambda$,
below which it confines with a chiral symmetry breaking.  A
holographic description of a strongly coupled cascading gauge theory
plasma is represented by a black brane solution of type IIB
supergravity on a conifold with fluxes.  A characteristic parameter
controlling the high temperature expansion of such plasma is
$\calq\simeq\left(\ln\frac T\Lambda\right)^{-1}$. In this paper we
study the speed of sound and the bulk viscosity of the cascading gauge
theory plasma to order $\calo(\calq^4)$. We find that the bulk
viscosity satisfies the bound conjectured in arXiv:0708.3459. We
comment on difficulties of computing the transport coefficients to all
orders in $\calq$.  Previously, it was shown that a cascading gauge
theory plasma undergoes a first-order deconfinement transition with
unbroken chiral symmetry at $T_{critical}=0.6141111(3)\Lambda$. We
show here that a deconfined chirally symmetric phase becomes
perturbatively unstable at $T_{unstable}=0.8749(0)T_{critical}$. Near
the unstable point the specific heat diverges as $c_V\sim |1-\frac
{T_{unstable}}{T}|^{-1/2}$.
}

\makepapertitle

\body

\version\versionno
\tableofcontents

\section{Introduction}

Consider $\caln=1$ four-dimensional supersymmetric $SU(K+P)\times SU(K)$
gauge theory with two chiral superfields $A_1, A_2$ in the $(K+P,\overline{K})$
representation, and two fields $B_1, B_2$ in the $(\overline{K+P},K)$.
This gauge theory has two gauge couplings $g_1, g_2$ associated with 
two gauge group factors,  and a quartic 
superpotential
\begin{equation}
W\sim \tr \left(A_i B_j A_kB_\ell\right)\e^{ik}\e^{j\ell}\,.
\end{equation}
When $P=0$ above theory flows in the infrared to a 
superconformal fixed point, commonly referred to as Klebanov-Witten (KW) 
theory \cite{kw}. At the IR fixed point KW gauge theory is 
strongly coupled --- the superconformal symmetry together with 
$SU(2)\times SU(2)\times U(1)$ global symmetry of the theory implies 
that anomalous dimensions of chiral superfields $\gamma(A_i)=\gamma(B_i)=-\frac 14$, \ie non-perturbatively large.

When $P\ne 0$, conformal invariance of the above $SU(K+P)\times SU(K)$
gauge theory is broken. It is useful to consider an effective description 
of this theory at energy scale $\mu$ with perturbative couplings
$g_i(\mu)\ll 1$. It is straightforward to evaluate NSVZ beta-functions for 
the gauge couplings. One finds that while the sum of the gauge couplings 
does not run
\begin{equation}
\frac{d}{d\ln\mu}\left(\frac{\pi}{g_s}\equiv \frac{4\pi}{g_1^2(\mu)}+\frac{4\pi}{g_2^2(\mu)}\right)=0\,,
\eqlabel{sum}
\end{equation}
the difference between the two couplings is  
\begin{equation}
\frac{4\pi}{g_2^2(\mu)}-\frac{4\pi}{g_1^2(\mu)}\sim P \ \left[3+2(1-\g_{ij})\right]\ \ln\frac{\mu}{\Lambda}\,,
\eqlabel{diff}
\end{equation}
where $\Lambda$  is the strong coupling scale of the theory and $\g_{ij}$ is an anomalous dimension of operators $\tr A_i B_j$.
Given \eqref{diff} and \eqref{sum} it is clear that the effective weakly coupled description of $SU(K+P)\times SU(K)$ gauge theory 
can be valid only in a finite-width energy band centered about $\mu$ scale. Indeed, extending effective description both to the UV 
and to the  IR one necessarily encounters strong coupling in one or the other gauge group factor. As beautifully explained 
in \cite{ks}, to extend the theory past the strongly coupled region(s) one must perform a Seiberg duality \cite{sd}. 
Turns out, in this gauge theory, a Seiberg duality transformation is a self-similarity transformation of the effective description 
so that $K\to K-P$ as one flows to the IR, or $K\to K+P$ as the energy increases. Thus, extension of the effective 
$SU(K+P)\times SU(K)$ description to all energy scales involves and infinite sequence - a {\it cascade } - of Seiberg dualities
where the rank of the gauge group is not constant along RG flow, but changes with energy according to \cite{b,k,aby2} 
\begin{equation}
K=K(\mu)\sim 2 P^2 \ln \frac \mu\Lambda\,, 
\eqlabel{effk}
\end{equation}
at least as $\mu\gg \Lambda$.
To see \eqref{effk}, note that the rank changes by $\Delta K\sim P$ as $P\Delta\left(\ln\frac\mu\Lambda\right)\sim 1$.
Although there are infinitely many duality cascade steps in the UV, there is only a finite number of duality transformations as one 
flows to the IR (from a given scale $\mu$). The space of vacua of a generic cascading gauge theory was studied in details in 
\cite{dks}. In the simplest case, when $K(\mu)$ is an integer multiple of $P$, the cascading gauge theory confines in the 
infrared with a spontaneous breaking of the chiral symmetry \cite{ks}. 

Effective description of the cascading gauge theory in the UV suggests that it must be ultimately defined as a theory with an infinite number 
of degree of freedom. If so, an immediate concern is whether such a theory is renormalizable as a four dimensional quantum field theory, \ie 
whether a definite prescription can be made for the computation of all gauge invariant correlation functions in the theory. 
As was pointed out in \cite{ks}, whenever $g_s K(\mu)\gg 1$, the cascading gauge theory allows for a dual holographic description \cite{juan,adscft} 
as type IIB 
supergravity on a warped deformed conifold with fluxes. The duality is always valid in the UV of the cascading gauge theory; if, in addition, 
$g_s P\gg 1$ the holographic correspondence is valid in the IR as well. It was shown in \cite{aby} that a cascading gauge theory {\it defined}
by its holographic dual  as an RG flow of type IIB supergravity on a warped deformed conifold with fluxes is holographically renormalizable 
as a four dimensional quantum field theory.  

Cascading gauge theories provide a soluble realization of the holographic gauge theory/string theory 
duality in non-conformal setting. A way to  construct four dimensional examples of non-conformal gauge theory/string theory correspondence 
is to start with an $AdS_5/CFT_4$ duality and to deform it by relevant operators of the $CFT_4$. An example of such construction is 
the gauge/string duality for the $\caln=2^*$ supersymmetric gauge theory \cite{pw,pwg1,pwg2}. On the contrary, the scale in a cascading 
gauge theory is  introduced via a dimensional transmutation of the gauge couplings \eqref{diff}. We would like to understand in details the 
hydrodynamic properties of strongly coupled non-conformal gauge theory plasmas \cite{ss}. A lot is known about thermodynamics/hydrodynamics of 
strongly coupled mass deformed conformal gauge theories from the perspective of gauge/string correspondence \cite{pwt1,pwt2,pwt3,pwt4,pwt5}.    
The thermodynamic/hydrodynamic analysis  of the cascading gauge theory plasma are substantially more difficult. The equilibrium thermodynamics 
of the cascading gauge theory plasma in the deconfined chirally symmetric phase is well understood by now \cite{kt1,kt2,kt3}. 
Since at zero temperature a cascading gauge theory confines with a chiral symmetry breaking, it is conceivable that there is a finite-temperature 
deconfined phase of the theory, with broken chiral symmetry\footnote{Such a phase was observed in $4+1$ dimensional supersymmetric $SU(N_c)$ 
gauge theory with fundamental quarks compactified on a circle \cite{asy}. }.  Whether or not such a phase exists is an 
open question \cite{wip}.   
In case of hydrodynamic transport coefficients\footnote{Hydrodynamics of closely related models was recently discussed in 
\cite{da}.}, the shear viscosity was shown to satisfy the universal bound \cite{u1,u2,u3}, 
and the bulk viscosity was computed to leading order at high temperature \cite{bb}. 
   
In this paper we study propagation of sound waves in the strongly coupled 
deconfined chirally symmetric phase of the cascading gauge theory plasma. In the dual 
gravitational description this involves computation of the dispersion relation of the lowest quasinormal mode 
in the sound channel \cite{ksq} of the black hole solution numerically constructed in \cite{kt3}.  
Ideally, we would like to do the analysis at any temperatures (at least above the deconfinement transition),
much like it was done for the $\caln=2^*$ plasma in \cite{pwt5}. Unfortunately, in section \ref{challenge}
we show that technical difficulties in the present framework does not allow us to achieve that goal. Thus, 
we resort to perturbative high temperature computations. The small parameter of the high temperature expansion is 
$\frac{P^2}{K_\star}$ \cite{kt2}, where   
$K_\star\simeq K(T)$ is roughly\footnote{A precise definition of $K_\star$ is given below.} 
the effective rank of the cascading plasma \eqref{effk} at the IR cutoff scale, set by the temperature.  
We compute both the speed of the sound waves and the bulk viscosity of the cascading gauge theory plasma to 
order $\calo\left(\frac{P^8}{K_\star^4}\right)$. Previously such analysis were done to order $\calo\left(\frac{P^2}{K_\star}\right)$ \cite{bb}.
At the order reported, the bulk viscosity was shown to saturate the bound proposed in \cite{bound}. 
Higher order corrections to the bulk viscosity of the cascading plasma presented here show that the bound 
\cite{bound} is satisfied. Alternative way to compute the speed of sound is to use the equilibrium 
equation of state
\begin{equation}
c_s^2=\frac{\del \calp}{\del \cale}\,.
\eqlabel{cs}
\end{equation} 
Using \eqref{cs} and slightly extending the computations in  \cite{kt3}, we can evaluate the speed of 
sound to temperatures down to the deconfinement transition and below. The comparison between the 
perturbative high temperature analysis and the exact one indicates that the former is convergent
for $\frac{K_\star}{P^2}\sim 2\cdots 3$, correspondingly to temperatures $T\simeq (1\cdots 1.5) \Lambda$ ---
which is about twice the critical temperature of the deconfinement phase transition 
$T_{critical}=0.6141111(3)\Lambda$ \cite{kt3}. Thus, although we find a relatively small 
bulk viscosity in the high temperature regime $\frac \zeta\eta \lesssim\frac 12$, we can not (reliably) evaluate 
the bulk viscosity in the vicinity of the deconfinement transition. It it clear though that 
since the deconfinement phase transition is of the  first-order (in the 't Hooft limit), 
the bulk viscosity will remain finite at the transition point \cite{bbss}.      
Finally, we find that chirally symmetric phase becomes perturbatively unstable at 
$T=T_{unstable}=0.87487(7) T_{critical}$ --- exactly at this temperature $c_s^2$ vanishes, 
and extending this phase to lower values of $K_{\star}$ leads to $c_s^2<0$. The critical behavior
at the unstable point in the cascading plasma is identical to the one found in $\caln=2^*$ plasma 
with mass deformation parameters $m_f< m_b$ \cite{bound,pwt5}. We comment more on 
the instability in section \ref{conclude}. 

The technical aspects of the computations are presented in section \ref{technical}.
The reader interested only in the results should consult section \ref{results}.

\section{Supergravity dual to deconfined cascading plasma }\label{technical}

The holographic dual to a deconfined chirally symmetric phase of a cascading gauge theory 
plasma at equilibrium is given by a black hole solution in a singular  Klebanov-Tseytlin (KT)
geometry \cite{kt}.  It has been discussed previously in \cite{b,kt1,kt2,kt3}. 
We follow the notations of \cite{kt3}.
A black hole metric\footnote{The frames $\{e_{\theta_a},e_{\phi_a}\}$ are defined as
in \cite{aby}, such that the metric on a unit size $T^{1,1}$ is
given by $\left(e_\psi^2\right)+ \sum_{a=1}^2
\left(e_{\theta_a}^2+e_{\phi_a}^2\right)$.} :
\begin{equation}
\begin{split}
ds_{10}^2=&h^{-1/2}(2x-x^2)^{-1/2}\left(-(1-x)^2 dt^2+dx_1^2+dx_2^2+dx_3^2\right)+G_{xx}(dx)^2\\
&+h^{1/2} [f_2\ \left(e_\psi^2\right)+ f_3\ \sum_{a=1}^2
\left(e_{\theta_a}^2+e_{\phi_a}^2\right)]\,,
\end{split}
\eqlabel{ktm}
\end{equation}
where $h$, $f_2$ and $f_3$ are some functions of the radial
coordinate $x$. There is also a dilaton $g(x)$, and form fields
given by
\begin{equation}
\begin{split}
&F_3=P\ e_\psi \wedge \left(e_{\theta_1}\wedge e_{\phi_1}-e_{\theta_2}\wedge e_{\phi_2}\right)\,,\qquad
B_2=\frac{K}{2 P}\ \left(e_{\theta_1}\wedge e_{\phi_1}-e_{\theta_2}\wedge e_{\phi_2}\right)\,,\\
&F_5=\calf_5+\star \calf_5\,,\qquad \calf_5=-K\ e_\psi\wedge e_{\theta_1}\wedge e_{\phi_1}\wedge e_{\theta_2}\wedge e_{\phi_2}\,,
\end{split}
\eqlabel{forms}
\end{equation}
where $K$ is a function of the radial coordinate $x$.  
Without loss of generality we can set
asymptotic string coupling to one. We use the following parametrization for the
solution in perturbation theory in $\frac{P^2}{K_\star}$ :
\begin{equation}
\begin{split}
&h(x)= \frac {K_{\star}}{4\tilde{a}_0^2}+\frac{K_\star}{\ta_0^2}\  
\sum_{n=1}^{\infty} \left\{\left(\frac{P^2}{K_\star}\right)^n
\left(\xi_{2n}(x)-\frac 54 \eta_{2n}(x)\right)\right\}\,,\\
\end{split}
\eqlabel{p2order1}
\end{equation}
\begin{equation}
\begin{split}
&f_2(x)=\tilde{a}_0+\ta_0\ \sum_{n=1}^\infty \left\{\left(\frac{P^2}{K_\star}\right)^n  \left(-2 \xi_{2n}(x)+\eta_{2n}(x)+\frac 45 \lambda_{2n}(x)\right)
\right\}\,,\\
\end{split}
\eqlabel{p2order2}
\end{equation}
\begin{equation}
\begin{split}
&f_3(x)=\tilde{a}_0+\ta_0\ \sum_{n=1}^\infty \left\{\left(\frac{P^2}{K_\star}\right)^n  \left(-2 \xi_{2n}(x)+\eta_{2n}(x)-\frac 15 \lambda_{2n}(x)\right)
\right\}\,,\\
\end{split}
\eqlabel{p2order3}
\end{equation}
\begin{equation}
\begin{split}
&K(x)=K_{\star}+K_{\star}\ \sum_{n=1}^\infty \left\{\left(\frac{P^2}{K_\star}\right)^n \k_{2n}(x)\right\}\,,\\
\end{split}
\eqlabel{p2order4}
\end{equation}
\begin{equation}
\begin{split}
&g(x)=1+\sum_{n=1}^\infty \left\{\left(\frac{P^2}{K_\star}\right)^n \zeta_{2n}(x)\right\}\,.\\
\end{split}
\eqlabel{p2order5}
\end{equation}
The advantage of this parametrization is that the equations for
$\{\xi_{2n},\eta_{2n},\lambda_{2n},\zeta_{2n}\}$ decouple, once the
(decoupled) equation for $\k_{2n}$ is solved, at each order
$n$ in perturbation theory.

Gauge invariant fluctuations $$
\{Z_H\,, Z_f\,, Z_\w\,, Z_\Phi\,, Z_K\}
$$
of the background metric \eqref{ktm} and the scalar fields 
$$\left\{f\equiv \frac 14 \ln h +\frac 25 \ln f_3+\frac {1}{10}\ln f_2\,,\qquad \w=\frac{1}{10}\ln\frac{f_3}{f_2}\,,\qquad \Phi=\ln g\,, 
\qquad K\right\}$$
of the effective five-dimensional gravitational description of the sound channel quasinormal modes were studied in details in \cite{bb}\footnote{See eq.~(63) of \cite{bb} for the definition 
of the gauge invariant fluctuations and eqs.~(64)-(68) of \cite{bb} for the corresponding 
equations of motion.}. 
The incoming wave boundary conditions on all physical modes imply that 
\begin{equation}
\begin{split}
&Z_H(x)=(1-x)^{-i\ww} z_H(x)\,,\qquad Z_f(x)=(1-x)^{-i\ww} z_f(x)\,,\\ 
&Z_w(x)=(1-x)^{-i\ww} z_w(x)\,, \qquad Z_\Phi(x)=(1-x)^{-i\ww} z_\Phi(x)\,,\\
&Z_K(x)=(1-x)^{-i\ww} z_K(x)\,,
\end{split}
\eqlabel{incoming}
\end{equation}
where $\{z_H, z_f, z_w,z_\Phi,z_K\}$ are regular at the horizon; we further introduced 
\begin{equation}
\ww\equiv \frac{\w}{2\pi T}\,,\qquad \qq\equiv\frac{q}{2\pi T} \,.
\eqlabel{defwwqq}
\end{equation}
$T$ is the equilibrium temperature of the plasma, and $\{\w,q=|\vec{ q}|\}$ are the 
frequency and the momentum of the 
sound quasinormal mode.
There is a single integration constant for  these  physical modes, namely, 
the overall scale. Without the loss of generality the latter can be fixed as 
\begin{equation}
z_H(x)\bigg|_{x\to 1_-}=1\,.
\eqlabel{bconditions}
\end{equation}
In this case, the sound dispersion relation is simply determined as
\begin{equation}
z_H(x)\bigg|_{x\to 0_+}=0\,.
\eqlabel{poledisp}
\end{equation}
The other boundary conditions (besides regularity at the horizon and \eqref{poledisp})
are 
\begin{equation}
z_f(x)\bigg|_{x\to 0_+}=0\,,\ z_w(x)\bigg|_{x\to 0_+}=0\,,\ z_\Phi(x)\bigg|_{x\to 0_+}=0\,,\ z_K(x)\bigg|_{x\to 0_+}=0\,.
\eqlabel{rembound}
\end{equation}
Let's  introduce 
\begin{equation}
\begin{split}
&z_H=\sum_{n=0}^{\infty} \biggl\{\left(\frac{P^2}{K_\star}\right)^{n}\ z_{H,0}^{(n)}\biggr\}+i\ \qq\  
\sum_{n=0}^{\infty} \biggl\{\left(\frac{P^2}{K_\star}\right)^{n}\ z_{H,1}^{(n)}\biggr\}\,,\\
&z_f=\sum_{n=0}^{\infty} \biggl\{\left(\frac{P^2}{K_\star}\right)^{n}\ z_{f,0}^{(n)}\biggr\}+i\ \qq\  
\sum_{n=0}^{\infty} \biggl\{\left(\frac{P^2}{K_\star}\right)^{n}\ z_{f,1}^{(n)}\biggr\}\,,\\
&z_\w=\sum_{n=0}^{\infty} \biggl\{\left(\frac{P^2}{K_\star}\right)^{n}\ z_{\w,0}^{(n)}\biggr\}+i\ \qq\  
\sum_{n=0}^{\infty} \biggl\{\left(\frac{P^2}{K_\star}\right)^{n}\ z_{\w,1}^{(n)}\biggr\}\,,\\
&z_\Phi=\sum_{n=0}^{\infty} \biggl\{\left(\frac{P^2}{K_\star}\right)^{n}\ z_{\Phi,0}^{(n)}\biggr\}+i\ \qq\  
\sum_{n=0}^{\infty} \biggl\{\left(\frac{P^2}{K_\star}\right)^{n}\ z_{\Phi,1}^{(n)}\biggr\}\,,\\
&z_K=K_\star \sum_{n=0}^{\infty} \biggl\{\left(\frac{P^2}{K_\star}\right)^{n}\ z_{K,0}^{(n)}\biggr\}+i\ \qq\ K_\star 
\sum_{n=0}^{\infty} \biggl\{\left(\frac{P^2}{K_\star}\right)^{n}\ z_{K,1}^{(n)}\biggr\}\,,
\end{split}
\eqlabel{defzz}
\end{equation}
where the lower index refers to either the leading, $\propto \qq^0$, or to the next-to-leading, $\propto \qq^1$,  
order in the hydrodynamic approximation, and the upper index keeps track of the $\frac{P^2}{K_\star}$ high temperature 
expansion parameter.   
Additionally, 
we find it convenient to parametrize 
\begin{equation}
\ww=\frac{\qq}{\sqrt{3}}\sum_{n=0}^\infty \biggl\{\left(\frac{P^2}{K_\star}\right)^{n} \b_{1,n} \biggr\}-i\   \frac {\qq^2}{3}
\sum_{n=0}^\infty \biggl\{\left(\frac{P^2}{K_\star}\right)^{n} \b_{2,n} \biggr\}\,,
\eqlabel{disprel}
\end{equation}
where  $\b_{1,n}$, $\b_{2,n}$ are constants which are to be determined from the pole dispersion relation
\eqref{poledisp}
\begin{equation}
\begin{split}
&z_{H,0}^{(n)}\bigg|_{x\to 0_+}=0\,,\qquad z_{H,1}^{(n)}\bigg|_{x\to 0_+}=0\,.
\end{split}
\eqlabel{poledisp1}
\end{equation}

\subsection{Equilibrium thermodynamics to order 
$\calo\left(\frac{P^8}{K_\star^4}\right)$}\label{sec21}

Cascading gauge theory thermodynamics for generic $\frac{P^2}{K_\star}$ was studied in \cite{kt3}. 
Since we are able to compute the $\calo\left(\qq^2\right)$ 
sound wave dispersion relation only perturbatively in $\frac{P^2}{K_\star}$, we need 
to construct the black hole backgrounds in the high temperature expansion as well. To order  
$\calo\left(\frac{P^2}{K_\star}\right)$ this was done in \cite{kt2}, and extended to order 
$\calo\left(\frac{P^6}{K_\star^3}\right)$ in \cite{kt3}. Here, we extend the analysis to $n=4$
in \eqref{p2order1}-\eqref{p2order5}.

Equations of motion for $\{\k_{2n}\,,\xi_{2n}\,,\eta_{2n}\,,\l_{2n}\,,\zeta_{2n}\}$ take form
\begin{equation}
0=\k_{2n}''+\frac{\k_{2n}'}{x-1}+\calj_{b,\k}^{[2n]}\,, \eqlabel{eqk2n}
\end{equation}
\begin{equation}
0=\eta_{2n}''+\frac{\eta_{2n}'}{x-1}-\frac{8\eta_{2n}}{x^2(x-2)^2}-\frac
25 \k_2'\ \k_{2n}'-\frac{8\k_{2n}}{3x^2(x-2)^2} +\calj_{b,\eta}^{[2n]}\,,
\eqlabel{eqeta2n}
\end{equation}
\begin{equation}
0=\xi_{2n}''+\frac{(3x^2-6x+4)\xi_{2n}'}{x(x-1)(x-2)}-\frac 23
\k_2'\ \k_{2n}'+\calj_{b,\xi}^{[2n]}\,, \eqlabel{eqxi2n}
\end{equation}
\begin{equation}
0=\lambda_{2n}''+\frac{\lambda_{2n}'}{x-1}-\frac{3\lambda_{2n}}{x^2(x-2)^2}-2
\k_2'\ \k_{2n}'+\calj_{b,\lambda}^{[2n]}\,, \eqlabel{eqlambda2n}
\end{equation}
\begin{equation}
0=\zeta_{2n}''+\frac{\zeta_{2n}'}{x-1}+2 \k_2'\
\k_{2n}'+\calj_{b,\zeta}^{[2n]}\,, \eqlabel{eqzeta2n}
\end{equation}
where the source terms
$\{\calj_{b,\k}^{[2n]},\calj_{b,\eta}^{[2n]},\calj_{b,\xi}^{[2n]},\calj_{b,\lambda}^{[2n]},\calj_{b,\zeta}^{[2n]}\}$
are functionals of the lower order solutions: $\k_{2m}$, $\xi_{2m}$,
$\eta_{2m}$, $\lambda_{2m}$, $\zeta_{2m}$, with $m<n$. Explicit
expressions for the source term functionals are available from the
author upon request.
The perturbative solutions to  \eqref{eqk2n}-\eqref{eqzeta2n} must
be regular at the horizon, and must have the appropriate KT
asymptotics  near the boundary. 
Beyond $n=1$, these equations
must be solved numerically. We apply the numerical strategy developed in \cite{kt3}:
\nxt Generically, the differential equations will have non-normalizable modes near the boundary\footnote{ These modes 
are  singular in case they are dual to operators of dimension larger than four ---  as for $\{f,\w\}$; 
in other cases they modify the KT asymptotics, \ie the parameters (such as a strong coupling scale) of the dual 
plasma. } $x\to 0_+$, and  can generate singular Schwarzschild horizon as $x\to 1_-$.   
Thus, we specify boundary conditions as a series expansion near the boundary and the horizon which explicitly 
contain only normalizable modes. 
\nxt The total number of integration constants near the boundary and the horizon appearing in  normalizable 
modes precisely equals the total order of the system of ODE's. As a result the boundary value problem is well posed.
\nxt We solve the resulting boundary value problem as detailed in section 5.2 of \cite{kt3}.

In what follows we present the horizon and the boundary expansion of the normalizable modes 
for $\{\k_{2n}\,,\xi_{2n}\,,\eta_{2n}\,,\l_{2n}\,,\zeta_{2n}\}$ with $n=\{1,2,3,4\}$. 
In numerical analysis we used expansion to order  $\calo(x^{9/2})$ (up to powers of $\ln x$)
near the boundary; and to order $\calo(y^{10})$, $y=1-x$, near the horizon. 
Below, however, we present expansion at most to orders $\calo(x^2)$ and $\calo(y^2)$ --- just what  is enough 
to exhibit the dependence on all the integration constants.
We solve the boundary value problem on the interval $x\in [\delta_x,1-\delta_x]$ with $\delta_x=10^{-2}$; 
we verified that our final results are insensitive to the precise choice of $\delta_x$, provided it is 
sufficiently small.
   
Notice from $\eqref{eqxi2n}$ that $\xi_{2n}$ always has a zero mode. Such a zero mode simply 
rescales (perturbatively in $\frac{P^2}{K_\star}$) $\ta_0$, and has no effect on physical quantities \cite{kt3}.
In what follows we conveniently set this mode to zero near the boundary\footnote{Of course, it is inconsistent 
to require the vanishing of the zero mode both near the boundary and the horizon. } --- for further details see 
\eqref{xo2}, \eqref{xo3}, \eqref{xo4} below. Naively,   from \eqref{eqzeta2n},  $\zeta_{2n}$ also 
always has a zero mode. The latter however is fixed by our choice of the asymptotic string coupling 
\eqref{p2order5}. Similarly, the zero mode of $\k_{2n}$, see  \eqref{eqk2n}, modifies the strong coupling scale 
$\Lambda$ of the cascading plasma.

\subsubsection{Order $n=1$}
 
We find:
\begin{equation}
\k_2=-\frac 12 \ln (2x-x^2)\,, \eqlabel{ko1}
\end{equation}
\begin{equation}
\xi_2=\frac{1}{12}\ln(2x-x^2)\,. \eqlabel{xio1}
\end{equation}
Even though it is possible to write down explicit analytic expressions for $\{\eta_2,\lambda_2,\zeta_2\}$,
such expressions involve complicated polylogarithm functions, which slows down subsequent numerical computations.
Thus, we opt to treat these fields numerically. 

Near the boundary, $x\to 0_+$,  we have:
\begin{equation}
\begin{split}
\eta_2=&-\frac 16+\frac 16 \ln 2+\frac 16 \ln x-\frac{1}{30} x+x^2 \left(\eta_{1}^{4,0}+\frac{1}{30} \ln x\right)+\calo(x^3\ \ln x)\,,
\end{split}
\eqlabel{eo1}
\end{equation}
\begin{equation}
\begin{split}
\l_2=&\frac 23 x+\l_1^{3,0} x^{3/2}+\calo(x^{2})\,,
\end{split}
\eqlabel{lo1}
\end{equation}
\begin{equation}
\begin{split}
\z_2=&x \left(\z_1^{2,0}+\frac 12 \ln x\right)+\calo(x^2\ \ln x)\,.
\end{split}
\eqlabel{zo1}
\end{equation}
Near the horizon, $y\to 0_+$,  we have:
\begin{equation}
\begin{split}
\eta_2=&\eta_{1,h}^{0}+\calo(y^2)\,,
\end{split}
\eqlabel{eo1h}
\end{equation}
\begin{equation}
\begin{split}
\l_2=&\l_{1,h}^0+\calo(y^2)\,,
\end{split}
\eqlabel{lo1h}
\end{equation}
\begin{equation}
\begin{split}
\z_2=&\z_{1,h}^0+\calo(y^2)\,.
\end{split}
\eqlabel{zo1h}
\end{equation}

Altogether at this order we have 6  integration constants
\begin{equation}
\{\eta_1^{4,0}\,, \l_1^{3,0}\,, \z_1^{2,0}\,, \eta_{1,h}^0\,, \l_{1,h}^0\,, \z_{1,h}^0\}\,,
\eqlabel{backorder1}
\end{equation}
which is precisely what is needed to specify a unique solution for $\{\eta_2,\l_2,\z_2\}$.

\subsubsection{Order $n=2$}

Near the boundary, $x\to 0_+$,  we have:
\begin{equation}
\begin{split}
\k_4=&x \left(\k_2^{2,0}-\frac 12 \ln x\right)+\calo(x^{3/2})\,,
\end{split}
\eqlabel{ko2}
\end{equation}
\begin{equation}
\begin{split}
\xi_4=&\frac{1}{36} \ln x+\calo(x \ \ln x)\,,
\end{split}
\eqlabel{xo2}
\end{equation}
\begin{equation}
\begin{split}
&\eta_4=-\frac{1}{12}+\frac{1}{18} \ln 2+\frac{1}{18} \ln x+x \left(-\frac{1}{30} \z_1^{2,0}+\frac{7}{360}-\frac{1}{60} \ln 2-\frac{7}{30} \k_2^{2,0}
+\frac{1}{12} \ln x\right)\\
&-\frac{4}{225} \l_1^{3,0} x^{3/2}+x^2 \biggl(\eta_{2}^{4,0}+\left(-\frac43 \eta_{1}^{4,0}+\frac{1}{18} \z_1^{2,0}-\frac{7}{1080}-\frac{1}{90}
 \ln 2+\frac{1}{15} \k_2^{2,0}\right) \ln x\\
&-\frac{11}{360} \ln^2 x\biggr)+\calo(x^{5/2})\,,
\end{split}
\eqlabel{eo2}
\end{equation}
\begin{equation}
\begin{split}
\l_4=&x \left(\frac 43 \k_2^{2,0}-\frac{11}{9}+\frac 23 \z_1^{2,0}+\frac 13 \ln 2\right)+x^{3/2} \l_2^{3,0}+\calo(x^{2}\ \ln x)\,,
\end{split}
\eqlabel{lo2}
\end{equation}
\begin{equation}
\begin{split}
\z_4=&x \left(\z_2^{2,0}+\left(\k_2^{2,0}-\frac 56+\frac 14 \ln 2+\frac 12 \z_1^{2,0}\right) \ln x\right)+\calo(x^{2}\ \ln^2 x)\,.
\end{split}
\eqlabel{zo2}
\end{equation}
Near the horizon, $y\to 0_+$,  we have:
\begin{equation}
\begin{split}
\k_4=&\k_{2,h}^0+\calo(y^2)\,,
\end{split}
\eqlabel{ko2h}
\end{equation}
\begin{equation}
\begin{split}
\xi_4=&\xi_{2,h}^0+\xi_{2,h}^1 y^2+\calo(y^4)\,,
\end{split}
\eqlabel{xo2h}
\end{equation}
\begin{equation}
\begin{split}
\eta_4=&\eta_{2,h}^0+\calo(y^2)\,,
\end{split}
\eqlabel{eo2h}
\end{equation}
\begin{equation}
\begin{split}
\l_4=&\l_{2,h}^0+\calo(y^2)\,,
\end{split}
\eqlabel{lo2h}
\end{equation}
\begin{equation}
\begin{split}
\z_4=&\z_{2,h}^0+\calo(y^2)\,.
\end{split}
\eqlabel{zo2h}
\end{equation}

Altogether at this order we have 10  integration constants
\begin{equation}
\{\k_2^{2,0}\,,\eta_2^{4,0}\,, \l_2^{3,0}\,, \z_2^{2,0}\,, \k_{2,h}^0\,,\xi_{2,h}^0\,, \xi_{2,h}^1\,,
\eta_{2,h}^0\,, \l_{2,h}^0\,, \z_{2,h}^0\}\,,
\eqlabel{backorder2}
\end{equation}
which is precisely what is needed to specify a unique solution for $\{\k_4,\xi_4,\eta_4,\l_4,\z_4\}$.

\subsubsection{Order $n=3$}

Near the boundary, $x\to 0_+$,  we have:
\begin{equation}
\begin{split}
\k_6=&x \left(\k_3^{2,0}+\left(-\k_2^{2,0}+\frac 56-\frac 14 \ln 2-\frac 12 \z_1^{2,0}\right) \ln x\right)+\calo(x^{3/2})\,,
\end{split}
\eqlabel{ko3}
\end{equation}
\begin{equation}
\begin{split}
\xi_6=&\left(-\frac{1}{108} \ln 2+\frac{1}{48}\right) \ln x+\calo(x\ \ln x)\,,
\end{split}
\eqlabel{xo3}
\end{equation}
\begin{equation}
\begin{split}
&\eta_6=-\frac{1}{54} \ln^2 2+\frac{17}{216} \ln 2-\frac{49}{648}+\left(-\frac{1}{54} \ln 2+\frac{1}{24}\right) \ln x+x \biggl(
-\frac{47}{1080}-\frac{1}{60} \z_1^{2,0}\ln 2 \\
&-\frac{1}{30} 
\k_2^{2,0} \ln 2+\frac{77}{2160} \ln 2+\frac{7}{360} \z_1^{2,0}-\frac{1}{60} \k_2^{2,0}-\frac{1}{30} \z_2^{2,0}-\frac{7}{30} \k_3^{2,0}
-\frac{1}{120} \ln^2 2
\\
&+\left(\frac 16 \k_2^{2,0}
+\frac{1}{12} \z_1^{2,0}+\frac{1}{24} \ln 2-\frac 19\right) \ln x\biggr)+x^{3/2} \left(-\frac{2}{225} \l_1^{3,0}-\frac{4}{225} \l_2^{3,0}\right)\\
&+x^2 \biggl(\eta_3^{4,0}+\biggl(\frac19 \z_1^{2,0} \k_2^{2,0}+\frac{1}{15} (\k_2^{2,0})^2
-\frac{1}{45} \k_2^{2,0} \ln 2+\frac{89}{72} \eta_{1}^{4,0}-\frac{19}{1440} \ln 2-\frac{163}{1620} \z_1^{2,0}\\
&-\frac{17}{90} \k_2^{2,0}
+\frac{25121}{259200}-\frac{17}{18} \eta_{1}^{4,0} \ln 2 
-\frac 43 \eta_{2}^{4,0}+\frac{37}{540} (\z_1^{2,0})^2+\frac{1}{18} \z_2^{2,0}+\frac{1}{15} \k_3^{2,0}\\
&-\frac{1}{180} \ln^2 2\biggr) \ln x+\left(-\frac{1}{15} \k_2^{2,0}+\frac{5}{12} \eta_{1}^{4,0}
-\frac{1}{30} \z_1^{2,0}-\frac{7}{720} \ln 2+\frac{421}{8640}\right) \ln^2 x\\\
&+\frac{1}{144} \ln^3 x\biggr)
+\calo(x^{5/2}\ \ln x)\,,
\end{split}
\eqlabel{eo3}
\end{equation}
\begin{equation}
\begin{split}
\l_6=&x \biggl(-\frac{61}{54} \ln 2+\frac{107}{54}+\frac 13  \z_1^{2,0}\ln 2+\frac 23 \k_2^{2,0} \ln 2+\frac 16 \ln^2 2-\frac{11}{9}
 \z_1^{2,0}-\frac{22}{9} \k_2^{2,0}+\frac 23 \z_2^{2,0}\\
&+\frac 43 \k_3^{2,0}\biggr)+x^{3/2} \l_3^{3,0}+\calo(x^2\ \ln x)\,,
\end{split}
\eqlabel{lo3}
\end{equation}
\begin{equation}
\begin{split}
\z_6=&x \biggl(\z_3^{2,0}+\biggl(-\frac 56 \z_1^{2,0}-\frac 53 \k_2^{2,0}+\frac 12 \z_2^{2,0}+\k_3^{2,0}+\frac{25}{18}
-\frac 56 \ln 2+\frac 18 \ln^2 2+\frac 14 \z_1^{2,0}\ln 2 \\
&+\frac 12 \k_2^{2,0} \ln 2\biggr) 
\ln x\biggr)+\calo(x^2\ \ln^2 x)\,.
\end{split}
\eqlabel{zo3}
\end{equation}
Near the horizon, $y\to 0_+$,  we have:
\begin{equation}
\begin{split}
\k_6=&\k_{3,h}^0+\calo(y^2)\,,
\end{split}
\eqlabel{ko3h}
\end{equation}
\begin{equation}
\begin{split}
\xi_6=&\xi_{3,h}^0+\xi_{3,h}^1 y^2+\calo(y^4)\,,
\end{split}
\eqlabel{xo3h}
\end{equation}
\begin{equation}
\begin{split}
\eta_6=&\eta_{3,h}^0+\calo(y^2)\,,
\end{split}
\eqlabel{eo3h}
\end{equation}
\begin{equation}
\begin{split}
\l_6=&\l_{3,h}^0+\calo(y^2)\,,
\end{split}
\eqlabel{lo3h}
\end{equation}
\begin{equation}
\begin{split}
\z_6=&\z_{3,h}^0+\calo(y^2)\,.
\end{split}
\eqlabel{zo3h}
\end{equation}

Altogether at this order we have 10 arbitrary integration constants
\begin{equation}
\{\k_3^{2,0}\,,\eta_3^{4,0}\,, \l_3^{3,0}\,, \z_3^{2,0}\,, \k_{3,h}^0\,,\xi_{3,h}^0\,, \xi_{3,h}^1\,,
\eta_{3,h}^0\,, \l_{3,h}^0\,, \z_{3,h}^0\}\,,
\eqlabel{backorder3}
\end{equation}
which is precisely what is needed to specify a unique solution for $\{\k_6,\xi_6,\eta_6,\l_6,\z_6\}$.

\subsubsection{Order $n=4$}
Ultimately, to evaluate the speed of sound and the bulk viscosity at this order we will need only 
$\k_8$ and $\xi_8$ solutions.

Near the boundary, $x\to 0_+$,  we have:
\begin{equation}
\begin{split}
\k_8=&x \biggl(\k_4^{2,0}+\biggl(\frac 56 \z_1^{2,0}+\frac 53 \k_2^{2,0}-\frac 12 \z_2^{2,0}-\k_3^{2,0}-\frac{25}{18}+\frac56 \ln 2-\frac 18 \ln^2 2
-\frac 14  \z_1^{2,0}\ln 2\\
&-\frac 12 
\k_2^{2,0} \ln 2\biggr) \ln x\biggr)+\calo(x^{3/2})\,,
\end{split}
\eqlabel{ko4}
\end{equation}
\begin{equation}
\begin{split}
\xi_8=&\left(\frac{1}{324} \ln^2 2-\frac{23}{1296} \ln 2+\frac{41}{1944}\right) \ln x+\calo(x\ \ln x)\,.
\end{split}
\eqlabel{xo4}
\end{equation}
Near the horizon, $y\to 0_+$,  we have:
\begin{equation}
\begin{split}
\k_8=&\k_{4,h}^0+\calo(y^2)\,,
\end{split}
\eqlabel{ko4h}
\end{equation}
\begin{equation}
\begin{split}
\xi_8=&\xi_{4,h}^0+\xi_{4,h}^1 y^2+\calo(y^4)\,.
\end{split}
\eqlabel{xo4h}
\end{equation}

Altogether at this order we have 4 integration constants
\begin{equation}
\{\k_4^{2,0}\,,\k_{4,h}^0\,,\xi_{4,h}^0\,,\xi_{4,h}^1\}\,,
\eqlabel{backorder4}
\end{equation}
which is precisely what is needed to specify a unique solution for $\{\k_8,\xi_8\}$.

\subsubsection{Integration constants for the normalizable modes}

Here we tabulate (see table~\ref{table1}) the integration constants for the normalizable modes of 
$\k_{2n}$, $\xi_{2n}$, $\eta_{2n}$, $\l_{2n}$, $\zeta_{2n}$ with $n=\{1,2,3,4\}$ obtained 
from solving the corresponding boundary value problems.
\bigskip

\begin{table}
\centerline{
\\
\begin{tabular}
{||c||c|c|c|c||}
	\hline
\textbf{\em n}  &  $1$    &   $2$  & $3$  & $4$\\
\hline
\hline
$\k_{n}^{2,0}$ &  &0.73675974 & -0.62226255&-0.03784377\\
$\eta_{n}^{4,0}$ &-0.01717287  &0.00534036 &-0.01064222 &\\
$\l_{n}^{3,0}$ & -0.87235794 &-1.11562943 & 1.39008636&\\
$\z_{n}^{2,0}$ & -0.15342641 &0.62226267 & -0.32514260&\\
$\k_{n,h}^{0}$ &  & 0.62226259& -0.42061461&0.00816831\\
$\xi_{n,h}^{0}$ &  &-0.07981931 & 0.01661150&-0.00920379\\
$\xi_{n,h}^{0}$ &  &0.01919989 & -0.05277626&0.01385333\\
$\eta_{n,h}^{0}$ &-0.14891337  &-0.21809464 & 0.00213345&\\
$\l_{n,h}^{0}$ &  0.16806881& -0.14619173& 0.01639579&\\
$\z_{n,h}^{0}$ & -0.41123352 &0.33024116 & -0.07445122&\\
\hline
\end{tabular}
}
\caption{Coefficients of the normalizable modes of the background geometry.
See \eqref{backorder1}, \eqref{backorder2}, \eqref{backorder3} and \eqref{backorder4}.}
\label{table1}
\end{table}

\subsubsection{$\calp$, $\cale$ and $c_s^2$ from equilibrium thermodynamics}

Using the results of \cite{kt3} we can compute 
\begin{equation}
\frac{\calp}{sT}=\frac 37\left(\frac{7}{12}-\ha_{2,0}\right)\,, \qquad 
\frac{\cale}{sT}=\frac 34 \left(1+\frac 47\ \ha_{2,0}\right)\,,
\eqlabel{pe}
\end{equation}
where $s$ is the entropy density\footnote{ 
Note that expressions in \eqref{pe} are valid for any temperature.}.
Furthermore, the perturbative high temperature expansion for 
$\ha_{2,0}$ is given by\footnote{In order to evaluate the coefficient of 
$\frac{P^8}{K_\star^4}$ term we need the boundary expansions for $\eta_4$ and $\lambda_4$ to order $\calo(x)$. These expansions do not 
depend on the coefficients of the corresponding normalizable modes $\{\eta_{4}^{4,0}\,, \l_4^{3,0}\}$.}  
\begin{equation}
\begin{split}
\ha_{2,0}=&\frac{7}{12} \frac{P^2}{K_\star}+\left(\frac 76 \k_2^{2,0}-\frac{35}{3}+\frac{7}{12} \z_1^{2,0}
+\frac{7}{24} \ln 2\right) \frac{P^4}{K_\star^2}
+\biggl(-\frac{35}{36} \ln 2-\frac{35}{18} \k_2^{2,0}+\frac{7}{48} \ln^2 2\\
&+\frac{175}{108}+\frac{7}{24} \z_1^{2,0}\ln 2 
+\frac{7}{12} \k_2^{2,0} \ln 2
+\frac{7}{12} \z_2^{2,0}+\frac 76 \k_3^{2,0}-\frac{35}{36} \z_1^{2,0}\biggr) \frac{P^6}{K_\star^3}+\biggl(
\frac{175}{72} \ln 2\\
&+\frac{7}{12} \z_3^{2,0}+\frac{175}{54} \k_2^{2,0}-\frac{35}{18} \k_2^{2,0} \ln 2-\frac{35}{36} \z_1^{2,0}\ln 2 
-\frac{875}{324}+\frac{175}{108} \z_1^{2,0}-\frac{35}{48} \ln^2 2\\
&+\frac{7}{96} \ln^3 2+\frac{7}{48}  \z_1^{2,0}\ln^2 2+\frac{7}{12} \k_3^{2,0}\ln 2 +\frac{7}{24}  \k_2^{2,0}\ln^2 2
+\frac{7}{24}  \z_2^{2,0}\ln 2+\frac 76 \k_4^{2,0}\\
&-\frac{35}{36} \z_2^{2,0}-\frac{35}{18} \k_3^{2,0}\biggr) 
\frac{P^8}{K_\star^4}+\calo\left(\frac{P^{10}}{K_\star^5}\right)\,.
\end{split}
\eqlabel{hadef}
\end{equation}
Note that the coefficient of $\frac{P^4}{K_\star^2}$ must vanish \cite{kt3} --- numerically 
we find that it is $\propto 2\times 10^{-10}$.

The precise temperature dependence of $K_\star$ was determined in \cite{kt3}
\begin{equation}
\frac{K_\star}{P^2}=\frac 12 \ln \left(\frac{64\pi^4 }{81}\ \times\ \frac{s T }{ \Lambda^4}\right)\,.
\eqlabel{ks}
\end{equation}
Using \eqref{ks} and  the expressions for the pressure and the energy density from  
\eqref{pe}  we find 
\begin{equation}
\begin{split}
c_s^2=\frac{\del\calp}{\del\cale}=\frac 13\ \frac{7-12\ha_{2,0}
-6 P^2\ \frac{d\ha_{2,0}}{d K\star}}{7+4 \ha_{2,0}+2 P^2\ \frac{d\ha_{2,0}}{d K\star}}\,,
\end{split}
\eqlabel{cs2gt}
\end{equation}
Thus, given the perturbative high temperature expansion for $\ha_{2,0}$ we can evaluate from 
\eqref{cs2gt} the 
perturbative high temperature expansion for $c_s^2$
\begin{equation}
\begin{split}
3 c_s^2=&=1-\frac 43 \frac{P^2}{K_\star}+\left(\frac{10}{3}
-\frac 23 \ln 2-\frac 83 \k_2^{2,0}-\frac 43 \z_1^{2,0}\right) \frac{P^4}{K_\star^2}+\biggl(
-8-\frac 23  \z_1^{2,0}\ln 2\\
&+\frac{10}{3} \ln 2-\frac 13 \ln^2 2-\frac 83 \k_3^{2,0}+\frac{40}{9} \z_1^{2,0}-\frac 43 \k_2^{2,0} \ln 2
+\frac{80}{9} \k_2^{2,0}-\frac 43 \z_2^{2,0}\biggr) \frac{P^6}{K_\star^3}\\
&+\biggl(-\frac 13 \ln^2 2 \z_1^{2,0}-\frac 43 \k_3^{2,0} \ln 2
-\frac 23 \k_2^{2,0}\ln^2 2 -\frac 23 \z_2^{2,0} \ln 2 +\frac{16}{9} (\k_2^{2,0})^2+\frac{16}{9} \k_2^{2,0} \z_1^{2,0}
\\
&+\frac 49 (\z_1^{2,0})^2-\frac 16 \ln^3 2-12 \ln 2+\frac{37}{9}  \z_1^{2,0}\ln 2+\frac{169}{9}+\frac{74}{9} \k_2^{2,0} \ln 2
+\frac 52 \ln^2 2\\
&-\frac{212}{9} \k_2^{2,0}-\frac{106}{9} \z_1^{2,0}+\frac{46}{9} \z_2^{2,0}+\frac{92}{9} \k_3^{2,0}
-\frac 43 \z_3^{2,0}-\frac 83 \k_4^{2,0}\biggr) \frac{P^8}{K_\star^4}+\calo\left(\frac{P^{10}}{K_\star^5}\right)\,.
\end{split}
\eqlabel{cs2eos}
\end{equation}
Notice that \eqref{cs2eos} provides {\it predictions} for $\b_{1,n}$ of \eqref{disprel} with $n=1,\cdots 4$. 

\subsection{Speed of sound waves in the cascading plasma to order 
$\calo\left(\frac{P^8}{K_\star^4}\right)$}

Equations of motion for the sound waves in the cascading plasma for generic $\frac{P^2}{K_\star}$ were derived
in \cite{bb}. Previously, they have been discussed (solved) only to order  $\calo\left(\frac{P^2}{K_\star}\right)$, 
\cite{bb}.
Here, we extend the analysis to $n=4$ in \eqref{defzz} at order $\calo(\qq^0)$. 

Equations of motion for $\{z_{H,0}^{(n)}\,, z_{f,0}^{(n)}\,, z_{\w,0}^{(n)}\,, z_{\Phi,0}^{(n)}\,, z_{K,0}^{(n)}\}$ take 
form\footnote{We used \eqref{ko1} and \eqref{xio1}.}
\begin{equation}
\begin{split}
0=&\left[z_{H,0}^{(n)}\right]''-\frac{3x^2-6x+2}{(x-1)(x^2-2x+2)}\ \left[z_{H,0}^{(n)}\right]'+
\frac{4}{x^2-2x+2}\ z_{H,0}^{(n)}\\
&+\frac{32}{x^2-2x+2}\ z_{K,0}^{(n)}-\frac{16x(2-x)}{3(1-x)}\ \k_{2n}'-\frac{16x^2(2-x)^2}{(x-1)(x^2-2x+2)}\ \xi_{2n}'
\\
&+\frac{8\b_{1,n}}{x^2-2x+2}+\calj_{s,H}^{[2n]}\,,
\end{split}
\eqlabel{eqzh1}
\end{equation}
\begin{equation}
\begin{split}
0=&\left[z_{f,0}^{(n)}\right]''+\frac{1}{x-1}\ \left[z_{f,0}^{(n)}\right]'-\frac{8}{x^2(2-x)^2}\ z_{f,0}^{(n)}
+\frac{3(x-1)}{10x(2-x)}\ \left[z_{K,0}^{(n)}\right]'\\
&+\frac{2}{x^2(2-x)^2}\ z_{K,0}^{(n)}+\frac{2}{3x(2-x)(1-x)^2}\ \k_{2n}+\calj_{s,f}^{[2n]}\,,
\end{split}
\eqlabel{eqzf1}
\end{equation}
\begin{equation}
\begin{split}
0=&\left[z_{\w,0}^{(n)}\right]''+\frac{1}{x-1}\ \left[z_{\w,0}^{(n)}\right]'-\frac{3}{x^2(2-x)^2}\ z_{\w,0}^{(n)}
+\frac{(x-1)}{5x(2-x)}\ \left[z_{K,0}^{(n)}\right]'\\
&+\frac{1}{15(1-x)^3}\ \l_{2n}'+\frac{1}{10x(2-x)(1-x)^2}\ \l_{2n}+\calj_{s,\w}^{[2n]}\,,
\end{split}
\eqlabel{eqzw1}
\end{equation}
\begin{equation}
\begin{split}
0=&\left[z_{\Phi,0}^{(n)}\right]''+\frac{1}{x-1}\ \left[z_{\Phi,0}^{(n)}\right]'
+\frac{2(x-1)}{x(2-x)}\ \left[z_{K,0}^{(n)}\right]'
+\frac{2}{3(x-1)^3}\ \z_{2n}'+\calj_{s,\Phi}^{[2n]}\,,
\end{split}
\eqlabel{eqzp1}
\end{equation}
\begin{equation}
\begin{split}
0=&\left[z_{K,0}^{(n)}\right]''+\frac{1}{x-1}\ \left[z_{K,0}^{(n)}\right]'
+\frac{2}{3(x-1)^3}\ \k_{2n}'+\calj_{s,K}^{[2n]}\,,
\end{split}
\eqlabel{eqzk1}
\end{equation}
where the source terms
$\{\calj_{s,H}^{[2n]},\calj_{s,f}^{[2n]},\calj_{s,\w}^{[2n]},\calj_{s,\Phi}^{[2n]},\calj_{s,K}^{[2n]}\}$
are functionals of the lower order solutions: $z_{H,0}^{(m)}$, $z_{f,0}^{(m)}$, $z_{\w,0}^{(m)}$, $z_{\Phi,0}^{(m)}$, $z_{K,0}^{(m)}$, 
$\k_{2m}$, $\xi_{2m}$,
$\eta_{2m}$, $\lambda_{2m}$, $\zeta_{2m}$ and $\b_{1,m}$, with $m<n$. Explicit
expressions for the source term functionals are available from the
author upon request. Apart from $n=0$  \cite{pss} and for $\{z_{K,0}^{(1)}\,,z_{H,0}^{(1)}\,, z_{\Phi,0}^{(1)} \}$
\cite{bb} these equations must be solved numerically. We use the same numerical approach as outlined in section \ref{sec21}.

\subsubsection{Order $n=0$}
We find:
\begin{equation}
z_{H,0}^{(0)}=2 x-x^2\,,\qquad z_{f,0}^{(0)}=z_{\w,0}^{(0)}=z_{\Phi,0}^{(0)}=z_{K,0}^{(0)}=0\,,
\eqlabel{sn01}
\end{equation}
\begin{equation}
\b_{1,0}=1\,.
\eqlabel{sn02}
\end{equation}

\subsubsection{Order $n=1$}
We find:
\begin{equation}
z_{K,0}^{(1)}=z_{H,0}^{(1)}=0\,,\qquad z_{\Phi,0}^{(1)}=\frac{x(2-x)}{12(1-x)^2}\ \ln(2x-x^2)\,,
\eqlabel{sn11}
\end{equation}
\begin{equation}
\b_{1,1}=-\frac 23\,.
\eqlabel{sn12}
\end{equation}

Near the boundary, $x\to 0_+$,  we have
\begin{equation}
z_{f,0}^{(1)}=-\frac{1}{80} x+x^2 \left(s_{f,1}^{4,0}-\frac{1}{60} \ln x\right)+\calo(x^3\ln x)\,,
\eqlabel{so11}
\end{equation}
\begin{equation}
z_{\w,0}^{(1)}=-\frac{1}{45} x+x^{3/2} s_{\w,1}^{3,0}+\calo(x^2)\,.
\eqlabel{so12}
\end{equation}
Near the horizon, $y\to 0_+$,  we have
\begin{equation}
z_{f,0}^{(1)}=s_{f,h}^{1,0}+\calo(y^2)\,,
\eqlabel{so11h}
\end{equation}
\begin{equation}
z_{\w,0}^{(1)}=s_{\w,h}^{1,0}+\calo(y^2)\,.
\eqlabel{so12h}
\end{equation}

Altogether at this order we have 4 integration constants
\begin{equation}
\{s_{f,1}^{4,0}\,,s_{\w,1}^{3,0}\,,s_{f,h}^{1,0}\,,s_{\w,h}^{1,0}\}\,,
\eqlabel{sorder1}
\end{equation}
which is precisely what is needed to specify a unique solution for $\{z_{f,0}^{(1)},z_{\w,0}^{(1)}\}$.

\subsubsection{Order $n=2$}

Near the boundary, $x\to 0_+$,  we have
\begin{equation}
z_{K,0}^{(2)}=x \left(s_{K,2}^{2,0}-\frac 16 \ln x\right)+\calo(x^{3/2})\,,
\eqlabel{so21}
\end{equation}
\begin{equation}
z_{H,0}^{(2)}=x\ s_{H,2}^{2,0}+\calo(x^{2}\ln x)\,,
\eqlabel{so25}
\end{equation}
\begin{equation}
\begin{split}
&z_{f,0}^{(2)}=x \left(\frac{13}{360}-\frac{1}{480} \ln 2+\frac{7}{40} s_{K,2}^{2,0}-\frac{1}{32} \ln x\right)
-\frac{2}{15} x^{3/2} s_{\w,1}^{3,0}+x^2 \biggl(s_{f,2}^{4,0}+\biggl(\frac{149}{4320}\\
&-\frac{11}{720} \ln 2-\frac{1}{20} s_{K,2}^{2,0}
-\frac{1}{60} \k_2^{2,0}
-\frac{1}{72} \z_{1}^{2,0}-\frac 12 s_{f,1}^{4,0}\biggr) \ln x+\frac{1}{720} \ln^2 x\biggr)+\calo(x^{5/2})\,,
\end{split}
\eqlabel{so22}
\end{equation}
\begin{equation}
z_{\w,0}^{(2)}=x \left(-\frac{2}{15} s_{K,2}^{2,0}+\frac{4}{45}-\frac{1}{45} \ln 2\right)+x^{3/2}\ s_{\w,2}^{3,0}+\calo(x^2\ln x)\,,
\eqlabel{so23}
\end{equation}
\begin{equation}
z_{\Phi,0}^{(2)}=x \left(s_{\Phi,2}^{2,0}+\left(-\frac 23+\frac 16 \ln 2+s_{K,2}^{2,0}\right) \ln x\right)+\calo(x^2\ln x)\,.
\eqlabel{so24}
\end{equation}
Near the horizon, $y\to 0_+$,  we have
\begin{equation}
z_{K,0}^{(2)}=s_{K,h}^{2,0}+\calo(y^2)\,,
\eqlabel{so21h}
\end{equation}
\begin{equation}
\begin{split}
&z_{H,0}^{(2)}=\biggl(-2 \b_{1,2}+320 s_{f,h}^{1,0}\ \eta_{1,h}^0-8 s_{K,h}^{2,0}-\frac 89-\frac 23 \z_{1,h}^0
+\frac 15 \l_{1,h}^0
-\frac{38}{3} \eta_{1,h}^0+\frac 12 (\l_{1,h}^0)^2+8 \xi_{2,h}^1
\\
&-8 s_{\w,h}^{1,0}+24 s_{w,h}^{1,0}\ \l_{1,h}^0
+\frac{56}{3} s_{f,h}^{1,0}\biggr) y^2
+\calo(y^4)\,,
\end{split}
\eqlabel{so22h}
\end{equation}
\begin{equation}
z_{f,0}^{(2)}=s_{f,h}^{2,0}+\calo(y^2)\,,
\eqlabel{so23h}
\end{equation}
\begin{equation}
z_{\w,0}^{(2)}=s_{\w,h}^{2,0}+\calo(y^2)\,,
\eqlabel{so24h}
\end{equation}
\begin{equation}
z_{\Phi,0}^{(2)}=s_{\Phi,h}^{2,0}+\calo(y^2)\,.
\eqlabel{so25h}
\end{equation}

Altogether at this order we have 10 integration constants
\begin{equation}
\{s_{K,2}^{2,0}\,,s_{H,2}^{2,0}\,,s_{f,2}^{4,0}\,,s_{\w,2}^{3,0}\,,s_{\Phi,2}^{2,0}\,,s_{K,h}^{2,0}
\,,\b_{1,2}\,,s_{f,h}^{2,0}\,,s_{\w,h}^{2,0}\,,s_{\Phi,h}^{2,0}\}\,,
\eqlabel{sorder2}
\end{equation}
which is precisely what is needed to specify a unique solution for 
$\{z_{K,0}^{(2)}$, $z_{H,0}^{(2)}$, $z_{f,0}^{(2)}$, $z_{\w,0}^{(2)}$, $z_{\Phi,0}^{(2)}\}$.

\subsubsection{Order $n=3$}

Near the boundary, $x\to 0_+$,  we have
\begin{equation}
z_{K,0}^{(3)}=x \left(s_{K,3}^{2,0}+\left(\frac{17}{24}-\frac 16 \ln 2-s_{K,2}^{2,0}\right) \ln x\right)+\calo(x^{3/2})\,,
\eqlabel{so31}
\end{equation}
\begin{equation}
z_{H,0}^{(3)}=x\ s_{H,3}^{2,0}+\calo(x^{2}\ln^2 x)\,,
\eqlabel{so35}
\end{equation}
\begin{equation}
\begin{split}
&z_{f,0}^{(3)}=x \biggl(-\frac{3}{80} \b_{1,2}+\frac{11}{480} \ln 2-\frac{1}{960} \ln^2 2-\frac{1}{12} s_{K,2}^{2,0}
-\frac{1}{160} s_{H,2}^{2,0}+\frac{7}{40} s_{K,3}^{2,0}
+\frac{1}{40} s_{\Phi,2}^{2,0}\\
&-\frac{11}{180}+\frac{7}{80} \ln 2\ s_{K,2}^{2,0}
+\biggl(\frac{83}{576}-\frac{1}{24} \ln 2-\frac{1}{16} s_{K,2}^{2,0}\biggr) \ln x-\frac{1}{64} \ln^2 x\biggr)
\\
&+x^{3/2} \biggl(\frac 19 s_{\w,1}^{3,0}-\frac{1}{15} s_{\w,1}^{3,0}\ \ln 2-\frac{2}{15} s_{\w,2}^{3,0}
-\frac{1}{15} s_{\w,1}^{3,0}\ \ln x\biggr)+x^2 \biggl(s_{f,3}^{4,0}+\biggl(
-\frac{7}{360} \ln 2\ \z_{1}^{2,0}\\
&-\frac{7}{120} \b_{1,2}+\frac{7}{24} s_{f,1}^{4,0}+\frac{67}{360} s_{K,2}^{2,0}
+\frac{239}{2160} \k_2^{2,0}+\frac{301}{4320} \z_{1}^{2,0}-\frac{1}{120} s_{H,2}^{2,0}+\frac{29}{960} \ln 2
-\frac 12 s_{f,2}^{4,0}\\
&-\frac{1}{20} s_{K,3}^{2,0}-\frac{1}{72} \z_{2}^{2,0}-\frac{1}{24} s_{\Phi,2}^{2,0}
-\frac{1}{60} \k_{3}^{2,0}-\frac{1}{12} \z_{1}^{2,0} s_{K,2}^{2,0}-\frac{1}{45} \k_2^{2,0} \ln 2-\frac{1}{10} \k_2^{2,0} s_{K,2}^{2,0}
\\
&-\frac 14 \ln 2\ s_{f,1}^{4,0}-\frac{1}{40} \ln 2\ s_{K,2}^{2,0}-\frac{11}{1440} \ln^2 2-\frac{6431}{103680}\biggr) \ln x
+\biggl(\frac{1}{480} \ln 2+\frac{1}{120} s_{K,2}^{2,0}\\
&+\frac{1}{720} \z_{1}^{2,0}+\frac{1}{360} \k_2^{2,0}
-\frac{461}{17280}\biggr) \ln^2 x\biggr)+\calo(x^{5/2}\ln x)\,,
\end{split}
\eqlabel{so32}
\end{equation}
\begin{equation}
\begin{split}
&z_{\w,0}^{(3)}=x \biggl(-\frac{1}{15} \b_{1,2}-\frac{89}{405}+\frac 29 s_{K,2}^{2,0}-\frac{1}{90} s_{H,2}^{2,0}-\frac{2}{15} s_{K,3}^{2,0}
-\frac{1}{15} s_{\Phi,2}^{2,0}+\frac{47}{540} \ln 2-\frac{1}{90} \ln^2 2
\\
&-\frac{1}{15} \ln 2\ s_{K,2}^{2,0}\biggr)+x^{3/2}\ s_{\w,3}^{3,0}+\calo(x^2\ln x)\,,
\end{split}
\eqlabel{so33}
\end{equation}
\begin{equation}
\begin{split}
&z_{\Phi,0}^{(3)}=x \biggl(s_{\Phi,3}^{2,0}+\biggl(
\frac 12 \b_{1,2}-\frac 53 s_{K,2}^{2,0}+\frac{1}{12} s_{H,2}^{2,0}+s_{K,3}^{2,0}+\frac 12 s_{\Phi,2}^{2,0}+\frac{1}{12} \ln^2 2
-\frac{47}{72} \ln 2\\
&+\frac 12 \ln 2\ s_{K,2}^{2,0}
+\frac{89}{54}\biggr) \ln x\biggr)+\calo(x^2\ln^2 x)\,.
\end{split}
\eqlabel{so34}
\end{equation}
Near the horizon, $y\to 0_+$,  we have
\begin{equation}
z_{K,0}^{(3)}=s_{K,h}^{3,0}+\calo(y^2)\,,
\eqlabel{so31h}
\end{equation}
\begin{equation}
z_{H,0}^{(3)}=\left(\frac 83 \b_{1,2}-2 \b_{1,3}+\cdots\right) y^2+\calo(y^4)\,,
\eqlabel{so32h}
\end{equation}
where $\cdots$ denote dependence on lower order coefficients, except for $\{\b_{1,2},\b_{1,3}\}$  --- the expression is too long to be presented here,
\begin{equation}
z_{f,0}^{(3)}=s_{f,h}^{3,0}+\calo(y^2)\,,
\eqlabel{so33h}
\end{equation}
\begin{equation}
z_{\w,0}^{(3)}=s_{\w,h}^{3,0}+\calo(y^2)\,,
\eqlabel{so34h}
\end{equation}
\begin{equation}
z_{\Phi,0}^{(3)}=s_{\Phi,h}^{3,0}+\calo(y^2)\,.
\eqlabel{so35h}
\end{equation}

Altogether at this order we have 10 integration constants
\begin{equation}
\{s_{K,3}^{2,0}\,,s_{H,3}^{2,0}\,,s_{f,3}^{4,0}\,,s_{\w,3}^{3,0}\,,s_{\Phi,3}^{2,0}\,,s_{K,h}^{3,0}\,,\b_{1,3}\,,s_{f,h}^{3,0}\,,s_{\w,h}^{3,0}
\,,s_{\Phi,h}^{3,0}\}\,,
\eqlabel{sorder3}
\end{equation}
which is precisely what is needed to specify a unique solution for 
$\{z_{K,0}^{(3)}$, $z_{H,0}^{(3)}$, $z_{f,0}^{(3)}$, $z_{\w,0}^{(3)}$, $z_{\Phi,0}^{(3)}\}$.

\subsubsection{Order $n=4$}

Near the boundary, $x\to 0_+$,  we have
\begin{equation}
\begin{split}
&z_{K,0}^{(4)}=x \biggl(s_{K,4}^{2,0}+\biggl(-\frac 12 \b_{1,2}-\frac{1}{12} \ln^2 2+\frac 74 s_{K,2}^{2,0}-\frac{1}{12} 
s_{H,2}^{2,0}-s_{K,3}^{2,0}-\frac 12 s_{\Phi,2}^{2,0}-\frac{265}{144}\\
&-\frac 12 \ln 2\ s_{K,2}^{2,0}
+\frac{17}{24} \ln 2\biggr) \ln x
+\frac{1}{48} \ln^2 x\biggr)+\calo(x^{3/2}\ln x)\,,
\end{split}
\eqlabel{so41}
\end{equation}
\begin{equation}
z_{H,0}^{(4)}=x\ s_{H,4}^{2,0}+\calo(x^{2}\ln^3 x)\,.
\eqlabel{so42}
\end{equation}
Near the horizon, $y\to 0_+$,  we have
\begin{equation}
z_{K,0}^{(4)}=s_{K,h}^{4,0}+\calo(y^2)\,,
\eqlabel{so41h}
\end{equation}
\begin{equation}
\begin{split}
&z_{H,0}^{(4)}=\biggl(\frac 83 \beta_{1,3}-2 \beta_{1,4}+\biggl(-\frac{14}{3}-48 s_{\w,h}^{1,0}\ \l_1^{h,0}-\frac{28}{3}
 \eta_1^{h,0}-640 s_{f,h}^{1,0}\ \eta_1^{h,0}-\frac{2}{15} \l_1^{h,0}\\
&+160 (\eta_1^{h,0})^2-\frac43 \z_1^{h,0}
-\frac{112}{3} s_{f,h}^{1,0} +16 s_{\w,h}^{1,0}+16 s_{K,h}^{2,0}+32 \xi_2^{h,1}+\frac{13}{5} (\l_1^{h,0})^2\biggr) 
\beta_{1,2}\\
&-5 \beta_{1,2}^2+\cdots\biggr) y^2+\calo(y^4)\,,
\end{split}
\eqlabel{so42h}
\end{equation}
where $\cdots$ denote dependence on lower order coefficients, except for $\{\b_{1,2},\b_{1,3},\b_{1,4}\}$  --- the expression is too long to be presented here.

Altogether at this order we have 4 integration constants
\begin{equation}
\{s_{K,4}^{2,0}\,,s_{H,4}^{2,0}\,,s_{K,h}^{4,0}\,,\b_{1,4}\}\,,
\eqlabel{sorder4}
\end{equation}
which is precisely what is needed to specify a unique solution for 
$\{z_{K,0}^{(4)},z_{H,0}^{(4)}\}$.

\subsubsection{Integration constants for the sound quasinormal  modes at $\calo(\qq^0)$}

Here we tabulate (see table~\ref{table2}) the integration constants for the normalizable modes of 
$\{z_{K,0}^{(n)}$, $z_{H,0}^{(n)}$, $z_{f,0}^{(n)}$, $z_{\w,0}^{(n)}$, $z_{\Phi,0}^{(n)}\}$. with $n=\{1,2,3,4\}$ obtained 
from solving the corresponding boundary value problems.
\bigskip

\begin{table}
\centerline{
\\
\begin{tabular}
{||c||c|c|c|c||}
	\hline
\textbf{\em n}  &  $1$    &   $2$  & $3$  & $4$\\
\hline
\hline
$s_{K,n}^{2,0}$ &  &0.07891997 & -0.53177623& 1.48077259\\
$s_{H,n}^{2,0}$  &  &-0.76150758 & 2.66710077& -5.15078326\\
$s_{f,n}^{4,0}$ &-0.01641302 & 0.02836594 & -0.00340767 &\\
$s_{\w,n}^{3,0}$  & 0.04361787 &-0.05326321 &-0.05682864 &\\
$s_{\Phi,n}^{2,0}$  & & -0.03656488 & -0.55146784 &\\
$s_{K,h}^{n,0}$  &  &0.14236466 & -0.61460152& 1.49213973\\
$\b_{1,n}$  &  & 0.33333333 & 0.340767665 & -0.81625254\\
$s_{f,h}^{n,0}$  & -0.00362335 & 0.038494905& -0.14888558&\\
$s_{\w,h}^{n,0}$  & -0.00413161 & 0.01711156&-0.04003014 &\\
$s_{\Phi,h}^{n,0}$  &  &0.33416570 &-0.74238359 &\\
\hline
\end{tabular}
}
\caption{Coefficients of the normalizable modes of the sound quasinormal modes to order  $\calo(\qq^0)$.
See \eqref{sorder1}, \eqref{sorder2}, \eqref{sorder3} and \eqref{sorder4}.}
\label{table2}
\end{table}

\subsection{Bulk viscosity of the cascading plasma to order 
$\calo\left(\frac{P^8}{K_\star^4}\right)$}

Equations of motion for the sound waves in the cascading plasma for generic $\frac{P^2}{K_\star}$ were derived
in \cite{bb}. Previously, they have been discussed (solved) only to order  $\calo\left(\frac{P^2}{K_\star}\right)$, 
\cite{bb}.
Here, we extend the analysis to $n=4$ in \eqref{defzz} at order $\calo(\qq^1)$. 

Equations of motion for $\{z_{H,1}^{(n)}\,, z_{f,1}^{(n)}\,, z_{\w,1}^{(n)}\,, z_{\Phi,1}^{(n)}\,, z_{K,1}^{(n)}\}$ take 
form\footnote{We used \eqref{ko1}, \eqref{xio1} and \eqref{sn01}, \eqref{sn02}.}
\begin{equation}
\begin{split}
0=&\left[z_{H,1}^{(n)}\right]''-\frac{3 x^2-6 x+2}{(x^2-2 x+2) (x-1)}\left[z_{H,1}^{(n)}\right]'+\frac{4}{x^2-2 x+2}z_{H,1}^{(n)}+\frac{32}{x^2-2 x+2}z_{K,1}^{(n)}
\\
&-\frac{2 x^2 (2-x)^2}{3^{1/2}(x-1) (x^2-2 x+2)^2} \left[z_{H,0}^{(n)}\right]'+\frac{4x (x-2) }{3^{1/2}(x^2-2 x+2)^2}z_{H,0}^{(n)}
\\
&+ \frac{64(2 x^2-4 x+3)}{3^{1/2}(x^2-2 x+2)^2} z_{K,0}^{(n)}
+\frac{16 (x^2-2 x+4) (2-x)^2 x^2}{3^{1/2}(x-1) (x^2-2 x+2)^2}\xi_{2n}'
+\frac{16 (x-2) x}{3^{3/2}(x-1)}\k_{2n}'\\
&-\frac{8\b_{2,n}}{3^{1/2}(x^2-2x+2)}-\frac{8 (x^2-2 x+4)\b_{1,n}}{3^{1/2}(x^2-2 x+2)^2}+\calj_{a,H}^{[2n]}\,,
\end{split}
\eqlabel{eqzh2}
\end{equation}
\begin{equation}
\begin{split}
0=&\left[z_{f,1}^{(n)}\right]''+\frac{1}{x-1}\left[z_{f,1}^{(n)}\right]'-\frac{8}{x^2(2-x)^2} z_{f,1}^{(n)}
-\frac{3(x-1)}{10x(x-2)} \left[z_{K,1}^{(n)}\right]'\\
&+\frac{2}{x^2(2-x)^2} z_{K,1}^{(n)}-\frac{2}{3^{1/2}(x-1)}\left[z_{f,0}^{(n)}\right]'+\frac{3^{1/2}}{10x(x-2)}
z_{K,0}^{(n)}\\
&+\frac{2}{3^{1/2}x(1-x)^2(x-2)}\k_{2n}
+\calj_{a,f}^{[2n]}\,,
\end{split}
\eqlabel{eqzf2}
\end{equation}
\begin{equation}
\begin{split}
0=&\left[z_{\w,1}^{(n)}\right]''+\frac{1}{x-1}\left[z_{\w,1}^{(n)}\right]'-\frac{3}{x^2(2-x)^2} z_{w,1}^{(n)}
-\frac{(x-1)}{5x(x-2)} \left[z_{K,1}^{(n)}\right]'\\
&-\frac{2}{3^{1/2}(x-1)}\left[z_{\w,0}^{(n)}\right]'+\frac{3^{1/2}}{15x(x-2)}
z_{K,0}^{(n)}+\frac{3^{1/2}}{15(x-1)^3} \l_{2n}'
\\
&+\frac{3^{1/2}}{10x(1-x)^2(x-2)}\l_{2n}+\calj_{a,\w}^{[2n]}\,,
\end{split}
\eqlabel{eqzw2}
\end{equation}
\begin{equation}
\begin{split}
0=&\left[z_{\Phi,1}^{(n)}\right]''+\frac{1}{x-1}\left[z_{\Phi,1}^{(n)}\right]'
-\frac{2(x-1)}{x(x-2)} \left[z_{K,1}^{(n)}\right]'
-\frac{2}{3^{1/2}(x-1)}\left[z_{\Phi,0}^{(n)}\right]'\\
&+\frac{2}{3^{1/2}x(x-2)}
z_{K,0}^{(n)}
-\frac{2}{3^{1/2}(x-1)^3}\z_{2n}'+\calj_{a,\Phi}^{[2n]}\,,
\end{split}
\eqlabel{eqzp2}
\end{equation}
\begin{equation}
\begin{split}
0=&\left[z_{K,1}^{(n)}\right]''+\frac{1}{x-1}\left[z_{K,1}^{(n)}\right]'
-\frac{2}{3^{1/2}(x-1)}\left[z_{K,0}^{(n)}\right]'
-\frac{2}{3^{1/2}(x-1)^3}\k_{2n}'+\calj_{a,K}^{[2n]}\,,
\end{split}
\eqlabel{eqzk2}
\end{equation}
where the source terms
$\{\calj_{a,H}^{[2n]},\calj_{a,f}^{[2n]},\calj_{a,\w}^{[2n]},\calj_{a,\Phi}^{[2n]},\calj_{a,K}^{[2n]}\}$
are functionals of the lower order solutions: $z_{H,1}^{(m)}$, $z_{f,1}^{(m)}$, $z_{\w,1}^{(m)}$, $z_{\Phi,1}^{(m)}$, $z_{K,1}^{(m)}$, 
 $z_{H,0}^{(m)}$, $z_{f,0}^{(m)}$, $z_{\w,0}^{(m)}$, $z_{\Phi,0}^{(m)}$, $z_{K,0}^{(m)}$,
$\k_{2m}$, $\xi_{2m}$,
$\eta_{2m}$, $\lambda_{2m}$, $\zeta_{2m}$ and $\{\b_{1,m},\b_{2,m}\}$, with $m<n$. Explicit
expressions for the source term functionals are available from the
author upon request. Apart from $n=0$  \cite{pss} and for $\{z_{K,1}^{(1)}\,,z_{H,1}^{(1)}\}$
\cite{bb} these equations must be solved numerically. We use the same numerical approach as outlined in section \ref{sec21}.

\subsubsection{Order $n=0$}
We find:
\begin{equation}
z_{H,1}^{(0)}=z_{f,1}^{(0)}=z_{\w,1}^{(0)}=z_{\Phi,1}^{(0)}=z_{K,1}^{(0)}=0\,,
\eqlabel{an01}
\end{equation}
\begin{equation}
\b_{2,0}=1\,.
\eqlabel{an02}
\end{equation}

\subsubsection{Order $n=1$}
We find:
\begin{equation}
z_{K,1}^{(1)}=z_{H,1}^{(1)}=0\,,
\eqlabel{an11}
\end{equation}
\begin{equation}
\b_{2,1}=\frac 23\,.
\eqlabel{an12}
\end{equation}

Near the boundary, $x\to 0_+$,  we have
\begin{equation}
z_{f,1}^{(1)}=\frac{3^{1/2}}{80} x+x^2 \left(a_{f,1}^{4,0}+\frac{7}{360} 3^{1/2} \ln x\right)+\calo(x^3\ln x)\,,
\eqlabel{ao11}
\end{equation}
\begin{equation}
z_{\w,1}^{(1)}=\frac{1}{45} 3^{1/2} x+x^{3/2}\ a_{\w,1}^{3,0}+\calo(x^2)\,,
\eqlabel{ao12}
\end{equation}
\begin{equation}
z_{\Phi,1}^{(1)}=x \left(a_{\Phi,1}^{2,0}-\frac 16 3^{1/2} \ln x\right)+\calo(x^2\ln x)\,.
\eqlabel{ao13}
\end{equation}
Near the horizon, $y\to 0_+$,  we have
\begin{equation}
z_{f,1}^{(1)}=a_{f,h}^{1,0}+\calo(y^2)\,,
\eqlabel{ao11h}
\end{equation}
\begin{equation}
z_{\w,1}^{(1)}=a_{\w,h}^{1,0}+\calo(y^2)\,,
\eqlabel{ao12h}
\end{equation}
\begin{equation}
z_{\Phi,1}^{(1)}=a_{\Phi,h}^{1,0}+\calo(y^2)\,.
\eqlabel{ao13h}
\end{equation}

Altogether at this order we have 6 integration constants
\begin{equation}
\{a_{f,1}^{4,0}\,,a_{\w,1}^{3,0}\,,a_{\Phi,1}^{2,0}\,,a_{f,h}^{1,0}\,,a_{\w,h}^{1,0}\,,a_{\Phi,h}^{1,0}\}\,,
\eqlabel{aorder1}
\end{equation}
which is precisely what is needed to specify a unique solution for $\{z_{f,1}^{(1)},z_{\w,1}^{(1)},z_{\Phi,1}^{(1)}\}$.

\subsubsection{Order $n=2$}

Near the boundary, $x\to 0_+$,  we have
\begin{equation}
z_{K,1}^{(2)}=x \left(a_{K,2}^{2,0}+\frac16 3^{1/2} \ln x\right)+\calo(x^{3/2})\,,
\eqlabel{ao21}
\end{equation}
\begin{equation}
z_{H,1}^{(2)}=x\ a_{H,2}^{2,0}+\calo(x^{2}\ln x)\,,
\eqlabel{ao25}
\end{equation}
\begin{equation}
\begin{split}
&z_{f,1}^{(2)}=x \left(-\frac{1}{36} 3^{1/2}+\frac{7}{40} a_{K,2}^{2,0}+\frac{1}{160} 3^{1/2} \ln 2
+\frac{1}{40} a_{\Phi,1}^{2,0}+\frac{1}{32} 3^{1/2} \ln x\right)-\frac{2}{15} x^{3/2}\ a_{\w,1}^{3,0}
\\
&+x^2 \biggl(a_{f,2}^{4,0}+\biggl(-\frac 12 a_{f,1}^{4,0}-\frac{1}{20} a_{K,2}^{2,0}-\frac{1}{40} 3^{1/2}
-\frac{1}{24} a_{\Phi,1}^{2,0}+\frac{7}{720} 3^{1/2} \ln 2+\frac{1}{72} 3^{1/2} \z_1^{2,0}\\
&+\frac{1}{60} 3^{1/2} 
\k_2^{2,0}\biggr) \ln x-\frac{1}{720} 3^{1/2} \ln^2 x\biggr)+\calo(x^{5/2})\,,
\end{split}
\eqlabel{ao22}
\end{equation}
\begin{equation}
z_{\w,1}^{(2)}=x \left(-\frac{2}{15} a_{K,2}^{2,0}-\frac{2}{27} 3^{1/2}-\frac{1}{15}
 a_{\Phi,1}^{2,0}+\frac{1}{90} 3^{1/2} \ln 2\right)+x^{3/2}\ a_{\w,2}^{3,0}+\calo(x^2\ln x)\,,
\eqlabel{ao23}
\end{equation}
\begin{equation}
z_{\Phi,1}^{(2)}=x \left(a_{\Phi,2}^{2,0}+\left(a_{K,2}^{2,0}+\frac 59 3^{1/2}+\frac 12 a_{\Phi,1}^{2,0}-\frac{1}{12} 3^{1/2} 
\ln 2\right) 
\ln x\right)+\calo(x^2\ln x)\,.
\eqlabel{ao24}
\end{equation}
Near the horizon, $y\to 0_+$,  we have
\begin{equation}
z_{K,1}^{(2)}=a_{K,h}^{2,0}+\calo(y^2)\,,
\eqlabel{ao21h}
\end{equation}
\begin{equation}
\begin{split}
&z_{H,1}^{(2)}=\biggl(-8 a_{K,h}^{2,0}+\frac{56}{3} 3^{1/2} s_{f,h}^{1,0}-8 3^{1/2} s_{\w,h}^{1,0}
+\frac{8}{45} 3^{1/2} \l_{1,h}^0-\frac{16}{3} 3^{1/2} \eta_{1,h}^0+\frac{17}{54} 3^{1/2}\\
&+320\ 3^{1/2} 
s_{f,h}^{1,0}\ \eta_{1,h}^0
+24\ 3^{1/2} s_{\w,h}^{1,0}\ \l_{1,h}^0-\frac{16}{3} 3^{1/2} \xi_{2,h}^1-8\ 3^{1/2} s_{K,h}^{2,0}
+\frac{56}{3} a_{f,h}^{1,0}-8 a_{\w,h}^{1,0}\\
&-2 a_{\Phi,h}^{1,0}+\frac 43 3^{1/2} \b_{1,2}+\frac 23 3^{1/2} \b_{2,2}-\frac{160}{3} 3^{1/2} 
(\eta_{1,h}^0)^2-\frac{8}{15} 3^{1/2} (\l_{1,h}^0)^2+320 a_{f,h}^{1,0}\ \eta_{1,h}^0\\
&+24 a_{\w,h}^{1,0}\
 \l_{1,h}^0\biggr) y^2
+\calo(y^4)\,,
\end{split}
\eqlabel{ao22h}
\end{equation}
\begin{equation}
z_{f,1}^{(2)}=a_{f,h}^{2,0}+\calo(y^2)\,,
\eqlabel{ao23h}
\end{equation}
\begin{equation}
z_{\w,1}^{(2)}=a_{\w,h}^{2,0}+\calo(y^2)\,,
\eqlabel{ao24h}
\end{equation}
\begin{equation}
z_{\Phi,1}^{(2)}=a_{\Phi,h}^{2,0}+\calo(y^2)\,.
\eqlabel{ao25h}
\end{equation}

Altogether at this order we have 10 integration constants
\begin{equation}
\{a_{K,2}^{2,0}\,,a_{H,2}^{2,0}\,,a_{f,2}^{4,0}\,,a_{\w,2}^{3,0}\,,a_{\Phi,2}^{2,0}\,,a_{K,h}^{2,0}
\,,\b_{2,2}\,,a_{f,h}^{2,0}\,,a_{\w,h}^{2,0}\,,a_{\Phi,h}^{2,0}\}\,,
\eqlabel{aorder2}
\end{equation}
which is precisely what is needed to specify a unique solution for 
$\{z_{K,1}^{(2)}$, $z_{H,1}^{(2)}$, $z_{f,1}^{(2)}$, $z_{\w,1}^{(2)}$, $z_{\Phi,1}^{(2)}\}$.

\subsubsection{Order $n=3$}

Near the boundary, $x\to 0_+$,  we have
\begin{equation}
z_{K,1}^{(3)}=x \left(a_{K,3}^{2,0}+\left(-\frac{43}{72} 3^{1/2}+\frac{1}{12} 3^{1/2} \ln 2
-a_{K,2}^{2,0}-\frac 12 a_{\Phi,1}^{2,0}\right) \ln x\right)+\calo(x^{3/2})\,,
\eqlabel{ao31}
\end{equation}
\begin{equation}
z_{H,1}^{(3)}=x\ a_{H,3}^{2,0}+\calo(x^{2}\ln^2 x)\,,
\eqlabel{ao35}
\end{equation}
\begin{equation}
\begin{split}
&z_{f,1}^{(3)}=x \biggl(\frac{49}{1080} 3^{1/2}-\frac{1}{160} a_{H,2}^{2,0}+\frac{1}{160} 3^{1/2} s_{H,2}^{2,0}
-\frac{7}{360} 3^{1/2} \ln 2+\frac{7}{80} \ln 2\ a_{K,2}^{2,0}\\
&+\frac{1}{320} 3^{1/2} \ln^2 2
+\frac{1}{40} a_{\Phi,2}^{2,0}+\frac{1}{80} \ln 2\ a_{\Phi,1}^{2,0}+\frac{7}{40} a_{K,3}^{2,0}
+\frac{3}{80} 3^{1/2} \b_{1,2}+\frac{1}{80} 3^{1/2} \b_{2,2}-\frac{1}{12} a_{K,2}^{2,0}\\
&-\frac{1}{240} a_{\Phi,1}^{2,0}
+\biggl(-\frac {71}{576} 3^{1/2}+\frac{1}{32} 3^{1/2} \ln 2-\frac{1}{16} a_{K,2}^{2,0}-\frac{1}{16} a_{\Phi,1}^{2,0}
\biggr) \ln x+\frac{1}{64} 3^{1/2} \ln^2 x\biggr)\\
&+x^{3/2} \biggl(\frac 19 a_{\w,1}^{3,0}-\frac{1}{15} a_{\w,1}^{3,0} \ln 2
-\frac{2}{15} a_{\w,2}^{3,0}-\frac{1}{15} a_{\w,1}^{3,0} \ln x\biggr)+x^2 \biggl(a_{f,3}^{4,0}+\biggl(\frac{3}{40} 3^{1/2} \b_{1,2}
\\
&+\frac{1}{360} 3^{1/2} s_{K,2}^{2,0}+\frac{7}{360} 3^{1/2} \b_{2,2}+\frac{7}{720} 3^{1/2} s_{H,2}^{2,0}
-\frac{77}{1440} 3^{1/2} \z_1^{2,0}-\frac{1}{360} 3^{1/2} \ln 2\\
&-\frac{43}{432} 3^{1/2} \k_2^{2,0}
+\frac{1}{120} 3^{1/2} \k_2^{2,0} \ln 2+\frac{1}{144} 3^{1/2} \ln 2 \z_1^{2,0}+\frac{11753}{518400} 3^{1/2}
+\frac{1}{72} 3^{1/2} \z_{2}^{2,0}\\
&+\frac{1}{60} 3^{1/2} \k_3^{2,0}+\frac{7}{1440} 3^{1/2} \ln^2 2
+\frac{161}{1440} a_{\Phi,1}^{2,0}+\frac{67}{360} a_{K,2}^{2,0}-\frac{1}{120} a_{H,2}^{2,0}-\frac{3}{40} \z_1^{2,0}\
 a_{\Phi,1}^{2,0}-\frac 12 a_{f,2}^{4,0}\\
&-\frac{1}{24} a_{\Phi,2}^{2,0}-\frac{1}{12} \z_1^{2,0}\ a_{K,2}^{2,0}-\frac{1}{12}
 \k_2^{2,0}\ a_{\Phi,1}^{2,0}-\frac{1}{20} a_{K,3}^{2,0}-\frac{1}{48} \ln 2\ a_{\Phi,1}^{2,0}-\frac{1}{10} \k_2^{2,0}\
 a_{K,2}^{2,0}\\
&-\frac{1}{40} \ln 2\ a_{K,2}^{2,0}-\frac 14 \ln 2\ a_{f,1}^{4,0}+\frac{7}{24} a_{f,1}^{4,0}\biggr) \ln x
+\biggl(-\frac{1}{720} 3^{1/2} \ln 2-\frac{1}{720} 3^{1/2} \z_1^{2,0}\\
&-\frac{1}{360} 3^{1/2} \k_2^{2,0}
+\frac{19}{640} 3^{1/2}+\frac{1}{240} a_{\Phi,1}^{2,0}+\frac{1}{120} a_{K,2}^{2,0}\biggr) \ln^2 x\biggr)+\calo(x^{5/2}\ln x)\,,
\end{split}
\eqlabel{ao32}
\end{equation}
\begin{equation}
\begin{split}
&z_{\w,1}^{(3)}=x \biggl(-\frac{1}{90} a_{H,2}^{2,0}-\frac{1}{15} a_{\Phi,2}^{2,0}-\frac{2}{15} a_{K,3}^{2,0}
-\frac{1}{30} \ln 2\ a_{\Phi,1}^{2,0}+\frac{71}{405} 3^{1/2}+\frac 29 a_{K,2}^{2,0}+\frac{11}{90} a_{\Phi,1}^{2,0}
\\
&+\frac{1}{90} 3^{1/2} s_{H,2}^{2,0}-\frac{1}{15} \ln 2\ a_{K,2}^{2,0}+\frac{1}{15} 3^{1/2} \b_{1,2}+\frac{1}{45} 3^{1/2} 
\b_{2,2}-\frac{8}{135} 3^{1/2} \ln 2+\frac{1}{180} 3^{1/2} \ln^2 2\biggr)\\
&+x^{3/2}\ a_{\w,3}^{3,0}+\calo(x^2\ln x)\,,
\end{split}
\eqlabel{ao33}
\end{equation}
\begin{equation}
\begin{split}
&z_{\Phi,1}^{(3)}=x \biggl(a_{\Phi,3}^{2,0}+\biggl(-\frac{1}{12} 3^{1/2} s_{H,2}^{2,0}+\frac{1}{12} a_{H,2}^{2,0}
+\frac 14 \ln 2\ a_{\Phi,1}^{2,0}+\frac 12 \ln 2\ a_{K,2}^{2,0}+\frac 12\ a_{\Phi,2}^{2,0}+a_{K,3}^{2,0}\\
&-\frac{71}{54} 3^{1/2}
-\frac 53 a_{K,2}^{2,0}-\frac{11}{12} a_{\Phi,1}^{2,0}-\frac{1}{2} 3^{1/2} \b_{1,2}-\frac 16 3^{1/2} \b_{2,2}
+\frac 49 3^{1/2} \ln 2-\frac{1}{24} 3^{1/2} \ln^2 2\biggr) \ln x\biggr)\\
&+\calo(x^2\ln^2 x)\,.
\end{split}
\eqlabel{ao34}
\end{equation}
Near the horizon, $y\to 0_+$,  we have
\begin{equation}
z_{K,1}^{(3)}=a_{K,h}^{3,0}+\calo(y^2)\,,
\eqlabel{ao31h}
\end{equation}
\begin{equation}
z_{H,1}^{(3)}=\biggl(-\frac 89 3^{1/2} \b_{2,2}+\frac 23 3^{1/2} \b_{2,3}+\cdots\biggr)y^2+\calo(y^4)\,,
\eqlabel{ao32h}
\end{equation}
where $\cdots$ denote dependence on lower order coefficients, except for $\{\b_{2,2},\b_{2,3}\}$  --- the expression is too long to be presented here,
\begin{equation}
z_{f,1}^{(3)}=a_{f,h}^{3,0}+\calo(y^2)\,,
\eqlabel{ao33h}
\end{equation}
\begin{equation}
z_{\w,1}^{(3)}=a_{\w,h}^{3,0}+\calo(y^2)\,,
\eqlabel{ao34h}
\end{equation}
\begin{equation}
z_{\Phi,1}^{(3)}=a_{\Phi,h}^{3,0}+\calo(y^2)\,.
\eqlabel{ao35h}
\end{equation}

Altogether at this order we have 10 integration constants
\begin{equation}
\{a_{K,3}^{2,0}\,,a_{H,3}^{2,0}\,,a_{f,3}^{4,0}\,,a_{\w,3}^{3,0}\,,a_{\Phi,3}^{2,0}\,,a_{K,h}^{3,0}\,,\b_{2,3}\,,a_{f,h}^{3,0}\,,a_{\w,h}^{3,0}
\,,a_{\Phi,h}^{3,0}\}\,,
\eqlabel{aorder3}
\end{equation}
which is precisely what is needed to specify a unique solution for 
$\{z_{K,1}^{(3)}$, $z_{H,1}^{(3)}$, $z_{f,1}^{(3)}$, $z_{\w,1}^{(3)}$, $z_{\Phi,1}^{(3)}\}$.

\subsubsection{Order $n=4$}

Near the boundary, $x\to 0_+$,  we have
\begin{equation}
\begin{split}
&z_{K,1}^{(4)}=x \biggl(a_{K,4}^{2,0}+\biggl(-\frac{1}{12} a_{H,2}^{2,0}-\frac 12 a_{\Phi,2}^{2,0}-a_{K,3}^{2,0}
-\frac{35}{72} 3^{1/2} \ln 2+\frac 74 a_{K,2}^{2,0}+a_{\Phi,1}^{2,0}+\frac 16 3^{1/2} \b_{2,2}\\
&+\frac{1}{24} 3^{1/2} \ln^2 2
+\frac{71}{48} 3^{1/2}+\frac{1}{12} 3^{1/2} s_{H,2}^{2,0}-\frac 12 \ln 2\ a_{K,2}^{2,0}-\frac 14 \ln 2\ a_{\Phi,1}^{2,0}
+\frac 12 3^{1/2} \b_{1,2}\biggr) \ln x\\
&-\frac{1}{48} 3^{1/2} \ln^2 x\biggr)+\calo(x^{3/2}\ln x)\,,
\end{split}
\eqlabel{ao41}
\end{equation}
\begin{equation}
z_{H,1}^{(4)}=x\ a_{H,4}^{2,0}+\calo(x^{2}\ln^3 x)\,.
\eqlabel{ao42}
\end{equation}
Near the horizon, $y\to 0_+$,  we have
\begin{equation}
z_{K,1}^{(4)}=a_{K,h}^{4,0}+\calo(y^2)\,,
\eqlabel{ao41h}
\end{equation}
\begin{equation}
\begin{split}
&z_{H,1}^{(4)}=\biggl(\frac 23 3^{1/2} \beta_{2,4}-\frac 89 3^{1/2} \b_{2,3}+\biggl(-\frac {16}{3} 3^{1/2} s_{\w,h}^{1,0}
-\frac{13}{15} 3^{1/2} (\l_{1,h}^0)^2
-\frac{160}{3} 3^{1/2} (\eta_{1,h}^0)^2\\
&-\frac{32}{3} 3^{1/2} \xi_{2,h}^1-\frac{16}{3} 3^{1/2} s_{K,h}^{2,0}
+\frac{10}{3} 3^{1/2} \b_{1,2}+\frac 49 3^{1/2} \z_{1,h}^0+\frac{640}{3} 3^{1/2} s_{f,h}^{1,0}\ \eta_{1,h}^0
\\
&+16\ 3^{1/2} s_{\w,h}^{1,0}\ \l_{1,h}^0+\frac{112}{9} 3^{1/2} s_{f,h}^{1,0}+\frac{28}{9} 3^{1/2} \eta_{1,h}^0
+\frac{2}{45} 3^{1/2} \l_{1,h}^0+\frac{14}{9} 3^{1/2}\biggr) \b_{2,2}+\cdots\biggr)y^2\\
&+\calo(y^4)\,,
\end{split}
\eqlabel{ao42h}
\end{equation}
where $\cdots$ denote dependence on lower order coefficients, except for $\{\b_{2,2},\b_{2,3},\b_{2,4}\}$  --- the expression is too long to be presented here.

Altogether at this order we have 4 integration constants
\begin{equation}
\{a_{K,4}^{2,0}\,,a_{H,4}^{2,0}\,,a_{K,h}^{4,0}\,,\b_{2,4}\}\,,
\eqlabel{aorder4}
\end{equation}
which is precisely what is needed to specify a unique solution for 
$\{z_{K,1}^{(4)},z_{H,1}^{(4)}\}$.

\subsubsection{Integration constants for the sound quasinormal  modes at $\calo(\qq^1)$}

Here we tabulate (see table~\ref{table3}) the integration constants for the normalizable modes of 
$\{z_{K,1}^{(n)}$, $z_{H,1}^{(n)}$, $z_{f,1}^{(n)}$, $z_{\w,1}^{(n)}$, $z_{\Phi,1}^{(n)}\}$. with $n=\{1,2,3,4\}$ obtained 
from solving the corresponding boundary value problems.
\bigskip

\begin{table}
\centerline{
\\
\begin{tabular}
{||c||c|c|c|c||}
	\hline
\textbf{\em n}  &  $1$    &   $2$  & $3$  & $4$\\
\hline
\hline
$a_{K,n}^{2,0}$ &  & 0.01069638& 0.23174494& -0.98113866\\
$a_{H,n}^{2,0}$  &  & 0.08509403& -0.78349123& 3.54269325\\
$a_{f,n}^{4,0}$  &  0.03737595& -0.06797689& 0.08231942&\\
$a_{\w,n}^{3,0}$  &  -0.08394261& 0.10167870& -0.05749855&\\
$a_{\Phi,n}^{2,0}$  & -0.29631916 & 0.38704391& -0.14363558&\\
$a_{K,h}^{n,0}$  &  & -0.18958164& 0.69624351&  -1.81005605\\
$\b_{2,n}$  &  & 0.13225837& -1.69770959 & 2.26988336\\
$a_{f,h}^{n,0}$  & 0.00580530 & -0.05087126& 0.17559889&\\
$a_{\w,h}^{n,0}$  &  0.00630707& -0.02271965& 0.05689077&\\
$a_{\Phi,h}^{n,0}$  & 0.11330816 & -0.38874716& 0.95684007&\\
\hline
\end{tabular}
}
\caption{Coefficients of the normalizable modes of the sound quasinormal modes to order  $\calo(\qq^1)$.
See \eqref{aorder1}, \eqref{aorder2}, \eqref{aorder3} and \eqref{aorder4}.}
\label{table3}
\end{table}

\section{Challenges of computing transport coefficients to all orders in 
$\frac{P^2}{K_\star}$ }\label{challenge}

In the previous section we detailed the computation of the speed of sound and the 
bulk viscosity of the cascading plasma, perturbatively in $\frac{P^2}{K_\star}$
(note that $\frac{P^2}{K_\star}\sim \left(\ln \frac T\Lambda\right)^{-1}$ 
for\footnote{For exact temperature dependence of $K_\star$ see \cite{kt3}.} $T\gg \Lambda$  \eqref{effk}. )
The results of the analysis presented in section \ref{results} indicate that although $\calo\left(\frac{P^8}{K_\star^4}\right)$
perturbative expansion is in excellent agreement with the full (nonperturbative in $\frac{P^2}{K_\star}$ computation for $c_s^2$
in the high temperature regime, this expansion does not converge below $T\simeq (1\cdots 1.5)\Lambda$, which is about twice as high 
as the temperature of the deconfinement phase transition. Thus, using perturbative analysis only we can not compute the 
bulk viscosity of the cascading plasma at the transition point. In this section we would like to explain the difficulty in 
going beyond the perturbative analysis as the latter might effect the analysis of other quasinormal modes in the 
cascading plasma, specifically those that could be responsible for the chiral symmetry breaking transition \cite{wip}.

To understand the problem, it is instructive to go back to the numerical computation of the cascading plasma equilibrium equation 
of state. This was solved both perturbatively  and non-perturbatively in $\frac{P^2}{K_\star}$ in \cite{kt3}. 
On the dual gravitational side this computation involves finding the black hole solution in asymptotic KT geometry, \ie 
determining the gravitational fields $\{h\,, f_2\,, f_3\,, K\,, g\}$ \eqref{p2order1}-\eqref{p2order5}. 
Above gravitational fields have non-normalizable (in some cases singular) modes both near the horizon $x\to 1_-$ 
and  near the boundary $x\to 0_+$. Thus numerical integration must be done on an open interval $x\in (0,1)$, \ie 
we need to provide the boundary conditions for the gravitational fields as the series expansion in $x$ near the boundary 
and in $y=1-x$ near the horizon. These series expansions must be fairly precise since, for examples the coefficient of the 
normalizable mode for $f_2$ (dual to the vev of the dimension eight gauge invariant operator of the cascading plasma) 
enters at order $x^2$ near the boundary, which is subdominant to coefficients $x^{n/2}\ln^k(x)$ with $n={0,1,2,3}$ 
of the general boundary expansion \cite{kt3} 
\begin{equation}
f_2=a_0+\sum_{n=1}^\infty\sum_{k=1}^{n} a_{n,k}\ x^{n/2}\ \ln^k x\,.
\eqlabel{f2expansion}
\end{equation}   
What saves the day, and ultimately allows for the full non-perturbative computations, is the fact that 
at each fixed order $n$, the maximum power of $\ln x$ in \eqref{f2expansion} happens to be bounded, 
$k\le n$. Thus, the series expansions of the type \eqref{f2expansion} are just generalized Taylor series 
expansions, which can be easily determined to any given order in $n$ --- the total number of expansion coefficients 
at order $n$ grows as $\calo(n^2)$. The situation would have been completely different, had the summation of 
$k$ extend to infinity. Here one would have to solve exactly for the  series in $\left(P^2 \ln x\right)$ at each order in 
$x$. Given the complexity of the equations involved the latter appears to be impossible. 

Unfortunately, precisely this problem occurs in computation of the quasinormal modes in the sound channel. 
Consider for example the gauge invariant fluctuation $z_f$ \eqref{incoming}. As for the gravitational field 
$f_2$, it depends on the vev of the dimension eight operator --- so its exact boundary asymptotic can not be specified with 
an accuracy of less than $\calo(x^{2})$\footnote{Of course, in order to get reliable numerical results boundary asymptotics 
must be more precise --- in our high temperature analysis we used expansions to order $\calo(x^{9/2})$, which is five more orders 
beyond the highest order at which the normalizable coefficients of the fluctuations enter.}. Collecting 
\eqref{so11}, \eqref{so22}, \eqref{so32} we find 
\begin{equation}
\begin{split}
z_{f,0}=&x\biggl( -\frac{1}{80}\frac{P^2}{K_\star}\ln^{0} x
+\frac{1}{720}\frac{P^4}{K_\star^2}\ln^1 x-\frac{1}{64}\frac{P^6}{K_\star^3}\ln^2 x+\cdots \biggr)
+\calo\left(\frac{P^{2k}}{K_\star^k}\ x^2 \ln^k x\right)\,,
\end{split}
\eqlabel{fullzf}
\end{equation}
where we explicitly indicated  only the leading $\ln x$ dependence at each order $\frac{P^{2k}}{K_\star^k}$.   
We further verified that $z_{f,0}^{(4)}=\calo\left(x\ \ln^3 x\right)$ as $x\to 0_+$, and that in fact all the perturbative expansions
for $\{z_H, z_f, z_\w, z_\Phi, z_K\}$ do not truncate in $k$ --- for example, 
\begin{equation}
z_{f,0}=\sum_{n=2}^\infty\ \sum_{k=0}^{\infty}\ s_f^{n,k}\  x^{n/2}\ \ln^k x \,. 
\eqlabel{nottruncate}
\end{equation}
It would be interesting to develop computational techniques to deal with this difficulty. 
Notice that the high temperature perturbative expansion provides an effective cutoff on the 
power of $\ln x$ in the boundary asymptotics since each additional factor of $\ln x$ comes with a factor of $\frac{P^2}{K_\star}$.

\section{Perturbative hydrodynamics of the cascading plasma}\label{results}

In this section we present results of the perturbative high temperature analysis of the speed of sound waves and the bulk viscosity in the 
cascading plasma. We begin with discussion of the consistency checks on our analysis. Next, we move towards discussion of comparison between 
exact speed of sound (as given by \eqref{cs2gt}) and its perturbative high temperature expansion. This will allow us to comment on the convergence properties of the 
high temperature expansion. Extending the numerical analysis of \cite{kt3} we show that chirally symmetric deconfined phase of the 
cascading gauge theory plasma becomes perturbatively unstable below the critical temperature of the deconfinement transition 
$T_{unstable}=0.8749(0)T_{critical}$. We comment on the possible source of the  instability. Finally, we discuss the bulk viscosity 
bound of \cite{bound} for the cascading plasma.  

\subsection{Consistency of analysis}

\subsubsection{The first law of thermodynamics}

Cascading gauge theory plasma has a single scale $\Lambda$. It only makes sense to discuss the thermodynamics/hydrodynamics of the theory, 
provided one keeps $\Lambda$ fixed. As explained in \cite{kt3}, enforcing that $\Lambda$ is temperature independent leads to the following 
condition 
\begin{equation}
\frac{d}{dT}\left(K_\star-\frac {P^2}{2} \ln a_0^2\right)=0\,,\qquad a_0\equiv \lim_{x\to 0_+} f_2\,.
\eqlabel{constrainta}
\end{equation}  
From \eqref{p2order2} we find
\begin{equation}
\begin{split}
a_0=&\ta_0\biggl(1-\frac 16 \frac{P^2}{K_\star}+\left(\frac{1}{18} \ln 2-\frac{1}{12}\right) \frac{P^4}{K_\star^2}-\left(
\frac{1}{54} \ln^2 2+\frac{49}{648}-\frac{17}{216} \ln 2\right) \frac{P^6}{K_\star^3}\\
&+\left(-\frac{55}{64}+\frac
{29}{243} \ln 2+\frac{1}{162} \ln^3 2-\frac{11}{216} \ln^2 2\right) \frac{P^8}{K_\star^4}+\calo\left(\frac{P^{10}}{K_\star^5}\right)\biggr)\,.
\end{split}
\eqlabel{a0a0t}
\end{equation}
Next, we  compute the temperature of the black hole \eqref{ktm} using \eqref{p2order1}-\eqref{p2order2}, and an explicit expression for 
$G_{xx}$ --- given by eq.~(2.6) of \cite{kt3}. We further invert the temperature relation to obtain $\ta_0$, and ultimately 
from \eqref{a0a0t} the perturbative expression 
for $a_0$:
\begin{equation}
\begin{split}
a_0=&\frac{T^2 \pi^2 K_\star}{4} \biggl(1+\frac 12 \frac{P^2}{K_\star}-\left(
-\frac{5}{12}+\frac 16 \ln 2 +15 (\eta_{1,h}^0)^2+6 \xi_{2,h}^0+\frac 25 (\l_{1,h}^0)^2\right) \frac{P^4}{K_\star^2}\\
&+\biggl(
\frac{95}{21}-\frac{25}{72} \ln 2 -3 \xi_{2,h}^0-\frac{15}{2} (\eta_{1,h}^0)^2
-\frac 15 (\l_{1,h}^0)^2-6 \xi_{3,h}^0+\frac{1}{18} \ln^2 2-\frac 45 \l_{1,h}^0\ \l_{2,h}^0\\
&+\frac 45 \eta_{1,h}^0\ (\l_{1,h}^0)^2
-30 \eta_{1,h}^0\ \eta_{2,h}^0
-40 (\eta_{1,h}^0)^3+\frac{4}{25} (\l_{1,h}^0)^3+30 \xi_{2,h}^0\ \eta_{1,h}^0\biggr) \frac{P^6}{K_\star^3}
\\
&+\calo\left(\frac{P^{8}}{K_\star^4}\right)\biggr)\,.
\end{split}
\eqlabel{a0tem}
\end{equation}
Using \eqref{a0tem} we find from \eqref{constrainta}
\begin{equation}
\begin{split}
\frac{dK_\star}{d\ln T}\bigg|_{d\Lambda=0}\equiv&\cala_2^{a}\ P^2+\cala_4^{a}\ \frac{P^4}{K_\star}+\cala_6^{a}\ \frac{P^6}{K_\star^2}
+\cala_8^{a}\ \frac{P^8}{K_\star^3}+\calo\left(\frac{P^{10}}{K_\star^4}\right)\\
=&2 P^2+ \frac{2 P^4}{ K_\star}+\frac{P^6}{ K_\star^2}+ \left(
-\frac76+\frac 23 \ln 2+60 (\eta_{1,h}^0)^2+24 \xi_{2,h}^0+\frac 85 (\l_{1,h}^0)^2\right) \frac{P^8}{K_\star^3 }\\
&+\calo\left(\frac{P^{10}}{K_\star^4}\right)\,.
\end{split}
\eqlabel{dkdt1}
\end{equation}

Given \eqref{pe}, \eqref{hadef}  and the expression for the entropy density $s$ 
of the black hole \eqref{ktm}, 
 the first law of thermodynamics 
$$
d\calp=s dT\,,
$$
leads to an alternative expression for $\frac{dK_\star}{d\ln T}$:
\begin{equation}
\begin{split}
&\frac{dK_\star}{d\ln T}\bigg|_{d\calp-s dT=0}\equiv\cala_2^{b}\ P^2+\cala_4^{b}\ 
\frac{P^4}{K_\star}+\cala_6^{b}\ \frac{P^6}{K_\star^2}
+\cala_8^{b}\ \frac{P^8}{K_\star^3}+\calo\left(\frac{P^{10}}{K_\star^4}\right)\\
=&2 P^2+\left(-\frac 43+2 \z_1^{2,0}+4 \k_2^{2,0}+\ln 2\right) \frac{P^4}{K_\star}
+\biggl(4 \k_3^{2,0}-\frac 83 \k_2^{2,0}-\frac 85 (\l_{1,h}^0)^2
-\frac43 \z_1^{2,0}+\frac{79}{18}\\
&-60 (\eta_{1,h}^0)^2+2 \z_2^{2,0}+2 \k_2^{2,0} \ln 2
+\frac 12 \ln^2 2-3 \ln 2
+\ln 2\ \z_1^{2,0}-24 \xi_{2,h}^0\biggr) \frac{P^6}{K_\star^2}
\\
&+\biggl(\frac{55}{9} \ln 2+\frac{91}{18} \z_1^{2,0}+\frac{91}{9} \k_2^{2,0}-\frac 73 \z_2^{2,0}-\frac {14}{3} \k_{3}^{2,0}
+2 \z_3^{2,0}+4 \k_4^{2,0}+16 \xi_{2,h}^0-\frac{7}{2} \ln 2\ \z_1^{2,0}\\
&-7 \k_2^{2,0} \ln 2-\frac 94 \ln^2 2+\ln 2\
 \z_2^{2,0}+\frac12 \ln^2 2\ \z_1^{2,0}+2 \ln 2\ \k_3^{2,0}+\ln^2 2\ \k_2^{2,0}+\frac14 \ln^3 2
\\
&+40 (\eta_{1,h}^0)^2+\frac{16}{15} (\l_{1,h}^0)^2-36 \xi_{3,h}^0-180 \eta_{1,h}^0\ \eta_{2,h}^0+\frac{24}{5} \eta_{1,h}^0\
 (\l_{1,h}^0)^2-\frac{24}{5} \l_{1,h}^0\ \l_{2,h}^0\\
&-240 (\eta_{1,h}^0)^3+\frac{24}{25} (\l_{1,h}^0)^3+180 \xi_{2,h}^0\ \eta_{1,h}^0
-30 \ln 2\ (\eta_{1,h}^0)^2-12 \ln 2\ \xi_{2,h}^0-\frac 45 \ln 2\ (\l_{1,h}^0)^2\\
&-\frac{607}{108}-120 (\eta_{1,h}^0)^2\ \k_2^{2,0}
-60 (\eta_{1,h}^0)^2\ \z_1^{2,0}-48 \xi_{2,h}^0\ \k_2^{2,0}-24 \xi_{2,h}^0\ \z_1^{2,0}-\frac{16}{5} (\l_{1,h}^0)^2\ \k_2^{2,0}
\\
&-\frac 85 (\l_{1,h}^0)^2\ \z_1^{2,0}\biggr) \frac{P^8}{K_\star^3}+\calo\left(\frac{P^{10}}{K_\star^4}\right)\,.
\end{split}
\eqlabel{dkdt2}
\end{equation}
Comparison between $\cala_{2n}^a$ and $\cala_{2n}^b$ provides a highly nontrivial test on the consistency of the analysis. 
Using the data from the table~\ref{table1}, 
the results of such comparison are presented in table~\ref{table4}.
 
\begin{table}
\centerline{
\\
\begin{tabular}
{||c||c|c|c|c||}
	\hline
\textbf{\em n}  &  $1$    &   $2$  & $3$  & $4$\\
\hline
\hline
$\biggl(1-\frac{\cala_{2n}^b}{\cala_{2n}^a}\biggr)$ & 0 & $8.2\times 10^{-9}$& $-3.5\times 10^{-7}$& $-6.1\times 10^{-7}$\\
\hline
\end{tabular}
}
\caption{Comparison between $\cala_{2n}^a$ and $\cala_{2n}^b$ of  \eqref{dkdt1} and \eqref{dkdt2}.}
\label{table4}
\end{table}

\subsubsection{$c_s^2$ from the equilibrium thermodynamics and from the hydrodynamics}

Our second consistency test compares the predictions of the equilibrium thermodynamics for $\b_{1,n}=\b_{1,n}^{thermo}$
from \eqref{cs2eos} with the direct computation of $\b_{1,n}=\b_{1,n}^{sound}$
(see \eqref{disprel} for the parametrization and table~\ref{table2} for the results)
. 
Using the data from the  table~\ref{table2}, 
the results of such comparison are presented in table~\ref{table5}.

\begin{table}
\centerline{
\\
\begin{tabular}
{||c||c|c|c|c||}
	\hline
\textbf{\em n}  &  $1$    &   $2$  & $3$  & $4$\\
\hline
\hline
$\biggl(1-\frac{\b_{1,n}^{thermo}}{\b_{1,n}^{sound}}\biggr)$ & 0 & $-1.8\times 10^{-8}$& $4.0\times 10^{-7}$& $5.5\times 10^{-7}$\\
\hline
\end{tabular}
}
\caption{Comparison of coefficients $\b_{1,n}$ in the sound wave 
dispersion relation \eqref{disprel} with the predicted values from the equilibrium thermodynamics, \eqref{cs2eos}.}
\label{table5}
\end{table}

\subsection{Speed of sound and perturbative instability of 
deconfined chirally symmetric phase of the cascading plasma at low temperatures}

\begin{figure}[t]
\begin{center}
\psfrag{cs2}{{$c_s^2$}}
\psfrag{ks}{\raisebox{3ex}{\footnotesize\hspace{-1.8cm}$k_s=\frac{K_\star}{P^2}-\frac 12\ln 2$}}
  \includegraphics[width=5in]{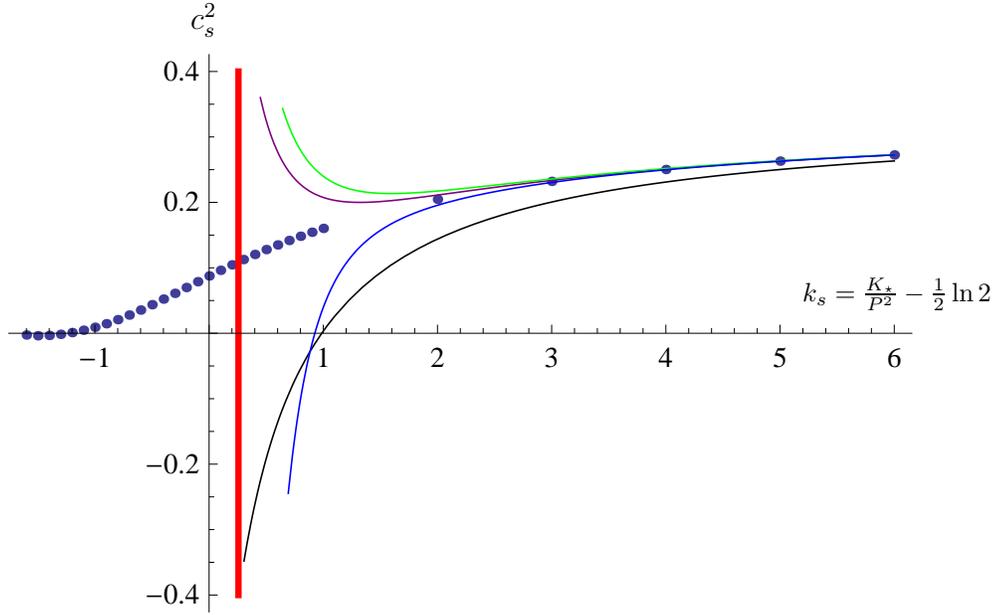}
\end{center}
  \caption{
(color online) Speed of sound (blue dots) vs. its successive high temperature approximations. The vertical red 
line represents the temperature of the deconfinement phase transition in the cascading plasma.  } \label{figure1}
\end{figure}

The speed of the sound waves can be computed from the dispersion relation of the quasinormal modes in the sound channel. 
In the high temperature expansion it is given by 
(see \eqref{disprel}) 
\begin{equation}
\begin{split}
c_s^2\bigg|_{high-T}=&\frac 13\biggl\{1-\frac43 \frac{P^2}{K_\star}+\left(2 \b_{1,2}+\frac 49\right) \frac{P^4}{K_\star^2}
+\left(2 \b_{1,3}-\frac 43 \b_{1,2}\right) \frac{P^6}{K_\star^3}\\
&+\left(2 \b_{1,4}-\frac 43 \b_{1,3}+(\b_{1,2})^2\right) \frac{P^8}{K_\star^4}+\calo\left(\frac{P^{10}}{K_\star^5}\right)\biggr\}\,,
\end{split}
\eqlabel{cs2ht}
\end{equation}
where the coefficients $\b_{1,n}$ are given in table~\ref{table2}.
Alternatively, it can be evaluated from the equilibrium thermodynamics for any temperature 
\eqref{cs2gt}, 
\begin{equation}
\begin{split}
c_s^2\bigg|_{thermo}=\frac{\del\calp}{\del\cale}=\frac 13\ \frac{7-12\ha_{2,0}
-6 P^2\ \frac{d\ha_{2,0}}{d K\star}}{7+4 \ha_{2,0}+2 P^2\ \frac{d\ha_{2,0}}{d K\star}}\,.
\end{split}
\eqlabel{cs2gt1}
\end{equation}
From data  in table~\ref{table5}, we see that there is an excellent agreement between \eqref{cs2ht} and the high temperature 
expansion of  \eqref{cs2gt}. Figure~\ref{figure1} represents comparison between \eqref{cs2gt} and \eqref{cs2ht} over a wide
range of temperatures, \ie not necessarily when $\frac{P^2}{K_\star}\ll 1$. The blue dots represent the speed of sound 
computed from \eqref{cs2gt}, slightly improving the analysis in \cite{kt3}. The
solid lines represent successive high temperature approximations to the speed of sound wave in the cascading plasma 
\eqref{cs2ht} to orders  $\calo\left(\frac{P^2}{K_\star}\right)$ (black), $\calo\left(\frac{P^4}{K_\star^2}\right)$ (purple), 
$\calo\left(\frac{P^6}{K_\star^3}\right)$ (green) and $\calo\left(\frac{P^8}{K_\star^4}\right)$ (blue).  
It is convenient to plot the data with respect to $k_s\equiv \frac{K_\star}{P^2}-\frac 12\ln 2$, rather than with respect to 
$\frac{T}{\Lambda}$.  The vertical red line represents the deconfinement temperature of the cascading gauge theory plasma. 
Notice that the second $\calo\left(\frac{P^4}{K_\star^2}\right)$ and the higher orders of the high temperature expansions 
become indistinguishable with the exact numerical data (blue dots) for $k_s\gtrsim 3 $.  Using results of \cite{kt3}
\begin{equation}
\begin{split}
&k_s=3\qquad \Longleftrightarrow\qquad \frac{T}{\Lambda}\approx 1.43\qquad \Longleftrightarrow\qquad 
\frac{T}{T_{critical}}\approx 2.34\,,\\
&k_s=2\qquad \Longleftrightarrow\qquad \frac{T}{\Lambda}\approx 1.00\qquad \Longleftrightarrow\qquad 
\frac{T}{T_{critical}}\approx 1.63\,,
\end{split}
\eqlabel{convergencet}
\end{equation} 
which suggests that the high temperature expansion  converges for temperatures above $T \gtrsim (1\cdots 1.5)\Lambda$.

\begin{figure}[t]
\begin{center}
\psfrag{cs2}{{$c_s^2$}}
\psfrag{tl}{{$\frac{T}{\Lambda}$}}
\psfrag{ks}{\raisebox{2ex}{\footnotesize\hspace{-0.5cm}$k_s$}}
  \includegraphics[width=3in]{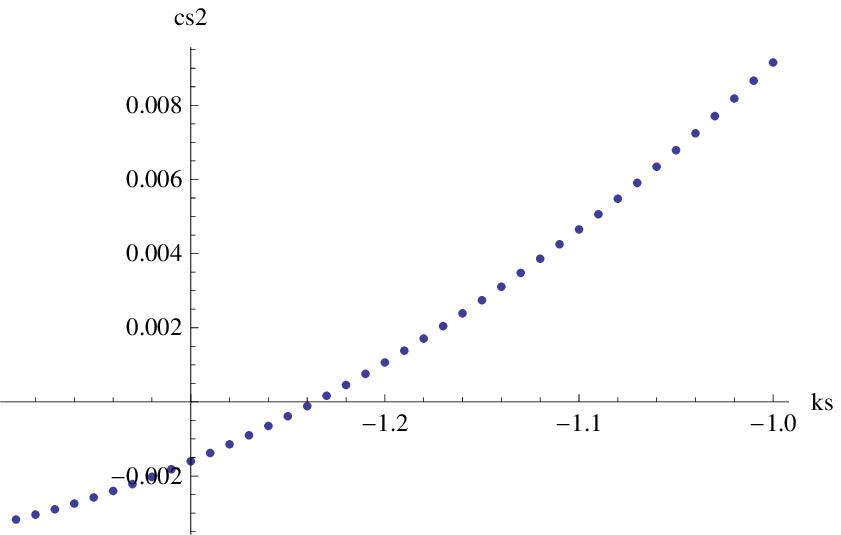}\
 \includegraphics[width=3in]{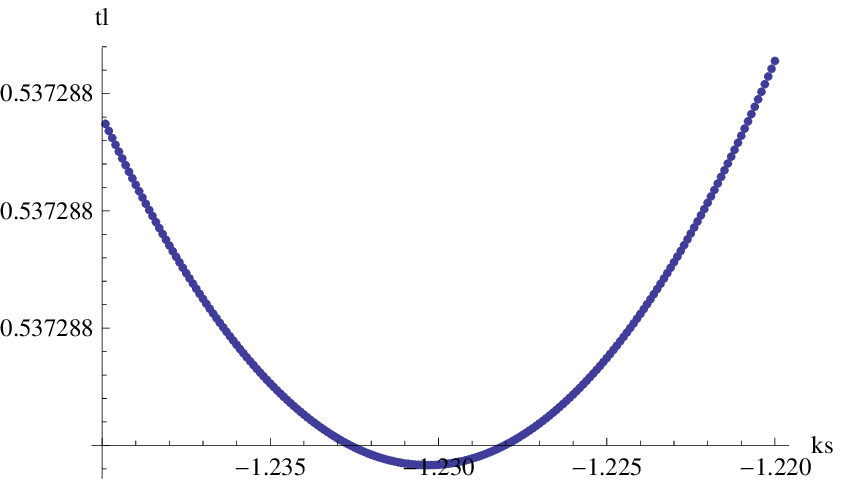}
\end{center}
  \caption{
Speed of sound in the vicinity of the perturbative instability of the cascading plasma. } \label{figure2}
\end{figure}

\begin{figure}[t]
\begin{center}
\psfrag{p}{{$\frac{32\pi^4}{81}\frac{\calf}{\Lambda^4}$}}
\psfrag{tl}{{$\frac{T}{\Lambda}$}}
\psfrag{ks}{\raisebox{2ex}{\footnotesize\hspace{-0.5cm}$k_s$}}
  \includegraphics[width=3in]{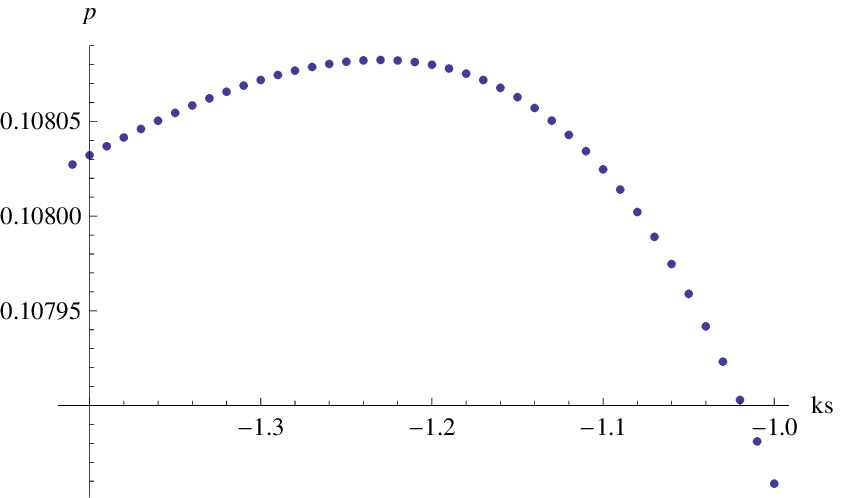}\
 \includegraphics[width=3in]{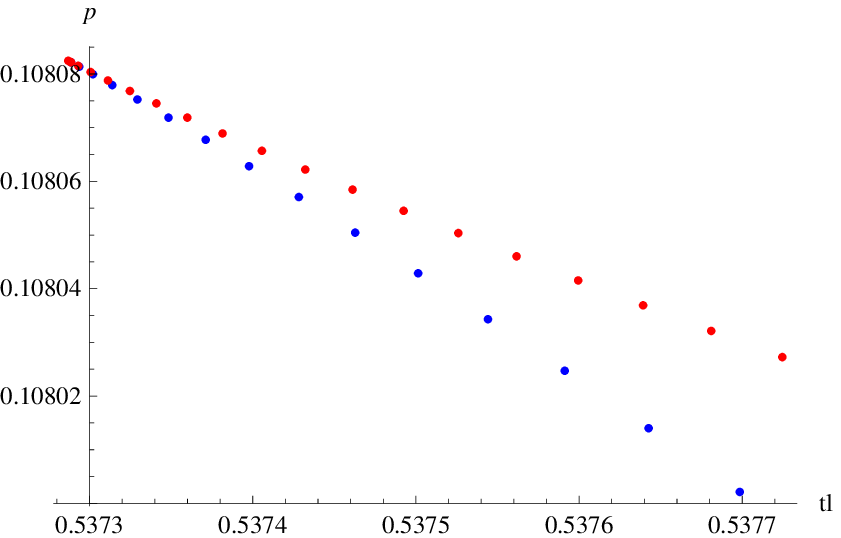}
\end{center}
  \caption{
(color online) 
The free energy density in the vicinity of the perturbative instability of the cascading plasma. The red dots on the right plot 
correspond to $k_s\le k_{unstable}$ and the blue dots correspond to $k_s\ge k_{unstable}$. } \label{figure3}
\end{figure}

Numerical analysis of the equilibrium thermodynamics in \cite{kt3} were done to temperatures only slightly
below the deconfinement temperature.  Here, we extend the computations to lower temperatures. 
Notice from figure~\ref{figure1} that the speed of sound squared $c_s^2$ appears to cross zero (and turns negative) 
for $k_s<-1$. A more detailed analysis presented in figure~\ref{figure2} show that this is indeed so.
A speed of sound vanishes at 
\begin{equation}
c_s^2(k_{unstable})=0\qquad \Longrightarrow\qquad k_{unstable}=-1.230(3)\,.
\eqlabel{kunst}
\end{equation}
While $c_s^2\sim (k_s-k_{unstable})$ in the vicinity of the instability,  we find
\begin{equation}
c_s^2\sim \left(1-\frac{T_{unstable}}{T}\right)^{1/2}\,,
\eqlabel{cs2cong}
\end{equation}
where 
\begin{equation}
\frac{T_{unstable}}{\Lambda}=0.53728(8)\,,\qquad {\rm or}\qquad T_{unstable}=0.8749(0)T_{critical}\,.
\eqlabel{tunst}
\end{equation} 
In fact, the near-unstable thermodynamics of the cascading theory is rather interesting. In figure~\ref{figure3} we show the free energy density
as a function of $k_s$ (left plot) and as a function of $\frac{T}{\Lambda}$ (right plot).  Notice that there are two phases separated by a continuous 
phase transition at $k_s=k_{unstable}$. The phase with $k_s>k_{unstable}$ (which is continuously connected to a high temperature 
deconfined chirally symmetric phase
of the cascading plasma) has a lower free energy compare to a phase with $k_s<k_{unstable}$ --- the latter two phases are degenerate in temperature 
(see the right plot on figure~\ref{figure2})
with the limiting temperature being reached precisely at $k_s=k_{unstable}$. It is clear from the right plot of figure~\ref{figure3} 
that\footnote{We did further, more detailed, numerical analysis to confirm this.} 
$$\frac{\del\calf}{\del T}\bigg|_{T\to T_{unstable}}$$
is finite. Thus, the only way the speed of sound can vanish at $T=T_{unstable}$ is if the rate of change of the energy density 
diverges at the unstable point. The plots in figure~4 show that this is indeed the case.

\begin{figure}[t]
\begin{center}
\psfrag{e}{{$\frac{32\pi^4}{81}\frac{\cale}{\Lambda^4}$}}
\psfrag{tl}{{$\frac{T}{\Lambda}$}}
\psfrag{ks}{\raisebox{2ex}{\footnotesize\hspace{-0.5cm}$k_s$}}
  \includegraphics[width=3in]{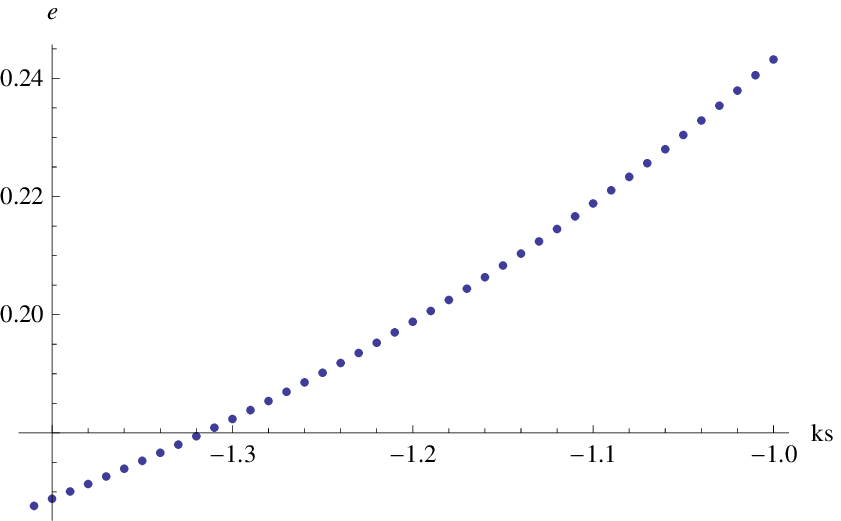}\
 \includegraphics[width=3in]{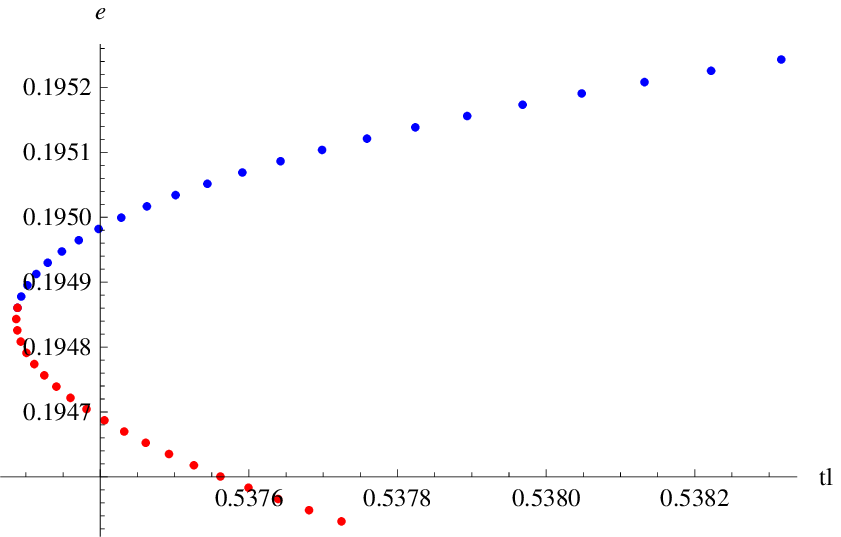}
\end{center}
  \caption{
(color online) 
The  energy density in the vicinity of the perturbative instability of the cascading plasma. The red dots on the right plot 
correspond to $k_s\le k_{unstable}$ and the blue dots correspond to $k_s\ge k_{unstable}$. } \label{figure4}
\end{figure}

Vanishing of the speed of sound as in \eqref{cs2cong} implies that the specific heat $c_V$ of the cascading plasma diverges near the unstable point 
with the critical exponent $\a=0.5$\footnote{This coincides with the mean-field critical exponent $\a$ at the tricritical point \cite{tc3}.}:
\begin{equation}
c_V=\frac{s}{c_s^2}\propto \bigg|1-\frac{T_{unstable}}{T}\bigg|^{-1/2}\,.
\eqlabel{cv}
\end{equation}
 Exactly the same critical behavior was found in $\caln=2^*$ plasma with mass deformation parameters 
$m_f<m_b$ \cite{bound,pwt5}.

Whenever $c_s^2<0$, the thermodynamic system is unstable with respect to density fluctuations. It is interesting to 
understand the source of this instability. 
Recall that the zero temperature the vacuum of the cascading gauge theory spontaneously breaks chiral symmetry.    
Thus, it is conceivable that the perturbative instability observed in the equilibrium thermodynamics of the 
deconfined chirally symmetric phase of the cascading plasma is associated with the formation of chiral 
condensates.  We comment more on this in the conclusion and for further analysis refer to future work \cite{wip}.

\subsection{Bulk viscosity bound in the cascading plasma}

\begin{figure}[t]
\begin{center}
\psfrag{z}{{$\left(\frac{\zeta}{\eta}-\delta\right)$}}
\psfrag{tl}{{$\frac{T}{\Lambda}$}}
\psfrag{d}{\raisebox{2ex}{\footnotesize\hspace{-0.5cm}$\delta$}}
  \includegraphics[width=5in]{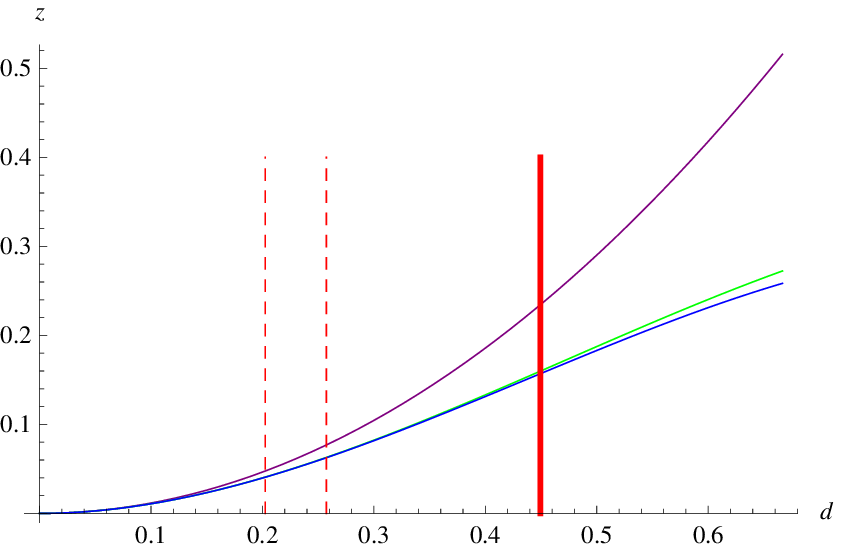}\
\end{center}
  \caption{
(color online) Successive high temperature approximations for the bulk viscosity bound for the cascading plasma. The viscosity 
bound implies $\frac{\zeta}{\eta}\ge \dd$. The solid vertical red line represents the temperature of the 
deconfinement phase transition in the cascading plasma. Vertical dashed lines indicated the expected 
convergence of the high temperature expansion: the left line corresponds to $T\approx 2.34 T_{critical}$
and the  right line corresponds to  $T\approx 1.63 T_{critical}$.}
\label{figure5}
\end{figure}

The primary goal of hydrodynamic analysis of the cascading plasma presented here was to verify the bulk viscosity 
bound in strongly coupled gauge theories conjectured in \cite{bound}:
\begin{equation}
\frac{\zeta}{\eta}\ge 2 \left(\frac 1p-c_s^2\right)\,,
\eqlabel{bbound}
\end{equation}
where $p$ is the dimension of space of the plasma. It is known that \eqref{bbound} is satisfied in all explicit realizations 
of gauge theory/string theory correspondence \cite{bound,pwt5}. The bound is saturated for all Dp branes \cite{bound,mas}. 
It is also known that one can engineer  phenomenological models motivated by gauge/string correspondence that would violate
the bulk viscosity bound \cite{gubser}; finally, the bound is violated in weakly coupled gauge theory plasmas \cite{adm}.

What makes cascading gauge theory plasma interesting in this context is that it provides an example in the framework of  the gauge/string 
duality where the bound \eqref{bbound} is saturated to leading order in the high temperature expansion \cite{bb}. As we explained in
section \ref{challenge} our current computational framework is inadequate to compute bulk viscosity of the cascading plasma at low temperatures. In the high temperature expansion it is given by (see \eqref{disprel})
\begin{equation}
\frac{\zeta}{\eta}=\frac 89\ \frac{P^2}{K_\star}+\frac 43\left(
\b_{2,2}\ \frac{P^4}{K_\star^2}+\b_{2,3}\ \frac{P^6}{K_\star^3}+\b_{2,4}\ \frac{P^8}{K_\star^4}
\right)+\calo\left(\frac{P^{10}}{K_\star^5}\right)\,.
\eqlabel{ze1}
\end{equation}
Using  \eqref{cs2ht} is can be rewritten as 
\begin{equation}
\begin{split}
\frac{\zeta}{\eta}=&
\dd+ \left( {\frac {27}{16}} {  \beta_{1,2}}+\frac 38+{\frac {27}{16}} {  
\beta_{2,2}} \right) {\dd}^{2}+ \biggl( {\frac {243}{128}} {  \beta_{1,3}}+{
\frac {729}{128}} {{  \beta_{1,2}}}^{2}+{\frac {81}{64}} {  \beta_{1,2}}+{
\frac {9}{32}}\\
&+{\frac {729}{128}} {  \beta_{2,2}} {  \beta_{1,2}}+{\frac {
81}{64}} {  \beta_{2,2}}+{\frac {243}{128}} {  \beta_{2,3}} \biggr) {\dd}^{3
}+ \biggl( {\frac {1215}{1024}} {  \beta_{1,2}}+{\frac {32805}{2048}} {
  \beta_{1,2}} {  \beta_{1,3}}+{\frac {98415}{4096}} {{  \beta_{1,2}}}^{3}\\
&+{
\frac {6561}{1024}} {{  \beta_{1,2}}}^{2}+{\frac {2187}{1024}} {  
\beta_{1,3}}+{\frac {135}{512}}+{\frac {2187}{1024}} {  \beta_{1,4}}+{\frac {
6561}{1024}} {  \beta_{2,2}} {  \beta_{1,3}}+{\frac {98415}{4096}} {  
\beta_{2,2}} {{  \beta_{1,2}}}^{2}\\
&+{\frac {6561}{1024}} {  \beta_{2,2}} {  
\beta_{1,2}}+{\frac {1215}{1024}} {  \beta_{2,2}}+{\frac {2187}{1024}} {  
\beta_{2,4}}+{\frac {19683}{2048}} {  \beta_{2,3}} {  \beta_{1,2}}+{\frac {2187
}{1024}} {  \beta_{2,3}} \biggr) {\dd}^{4}\\
&+\calo\left(\dd^5\right)\,,
\end{split}
\eqlabel{ze2}
\end{equation}
where we introduced 
\begin{equation}
\delta\equiv 2 \left(\frac 13-c_s^2\right)\,.
\eqlabel{defdelta}
\end{equation}
The results of the high temperature computations are presented in figure~\ref{figure5}. 
The
solid lines represent successive high temperature approximations to the bulk viscosity bound  in the cascading plasma \eqref{ze2} 
 to orders   $\calo\left(\frac{P^4}{K_\star^2}\right)$ (purple), 
$\calo\left(\frac{P^6}{K_\star^3}\right)$ (green) and $\calo\left(\frac{P^8}{K_\star^4}\right)$ (blue). Recall that the bound is exactly saturated at order 
 $\calo\left(\frac{P^2}{K_\star}\right)$. The vertical solid red line represents the value of $\dd=\dd_{critical}$ at the deconfinement phase transition.
The dashed red lines indicate the expected convergence of the high temperature expansion of the hydrodynamic quantities deduced from the high temperature 
expansion of the speed of sound, see figure~\ref{figure1}.

From figure.\ref{figure5} we  see that at least in the high temperature expansion the bulk viscosity bound \eqref{bbound} is satisfied. 
Since the deconfinement phase transition in the cascading plasma is of the first order \cite{kt3},  we do not expect any singular behavior in the 
bulk viscosity \cite{bbss} in the vicinity of the transition. Notice that there is almost no difference between $\calo\left(\frac{P^6}{K_\star^3}\right)$ (green) 
and $\calo\left(\frac{P^8}{K_\star^4}\right)$ (blue) approximations to the viscosity bound all the way to the deconfinement temperature $T_{critical}$.
This suggests that the high temperature expansion for the bulk viscosity might have better convergence properties than that of the speed of sound. 
If we take the high temperature results at the deconfinement transition seriously, we find that 
\begin{equation}
\frac{\zeta}{\eta}\bigg|_{deconfinement}\simeq 0.6(1)\,.
\eqlabel{zdeconf}
\end{equation}

QCD slightly above the deconfinement phase transition is nearly conformal. For $c_s^2$ in the range $0.27-0.31$, as in QCD at $T=1.5 T_{deconfinement}$  \cite{qcd1,qcd2}, 
we are well inside the expected validity range of the high temperature expansion, resulting in    
\begin{equation}
\frac{\zeta}{\eta}\bigg|_{QGP}\approx 0.05 - 0.14
\eqlabel{qgp}
\end{equation}
for the cascading gauge theory plasma.

\section{Conclusion}\label{conclude}
In this paper we presented detailed analysis of the transport properties of the deconfined chirally symmetric phase 
of the cascading plasma at strong coupling, using the gauge theory/string theory correspondence.
We developed the high temperature expansion to order $\calq^4\simeq \left(\ln \frac{T}{\Lambda}\right)^{-4}$ 
for the thermodynamic and the hydrodynamic properties of the theory and identified challenges in going beyond 
perturbative in $\calq$ hydrodynamics.  

We computed the high temperature expansion of the bulk viscosity of the cascading plasma. We showed that the bulk viscosity bound 
proposed in \cite{bound} is satisfied in such plasma. We argued that results for the bulk viscosity are likely to be reliable 
up to the deconfinement temperature with bulk viscosity being about $60\%$ of the shear viscosity 
right at the deconfinement transition. Much like in other holographic models of gauge theory/string theory duality \cite{pwt5}
we observe a rapid drop in the bulk viscosity above  the deconfinement transition.  

An interesting byproduct of our hydrodynamic analysis was the discovery of the perturbative instability 
of the deconfined chirally symmetric phase of the cascading plasma. Specifically, extending analysis of 
\cite{kt3} we identified a continuous phase transition in the plasma (slightly below the critical temperature 
of the first order deconfinement transition) where the speed of sound squared vanishes and becomes negative. 
A similar phase transition was observed previously in $\caln=2^*$ gauge theory plasma \cite{bound,pwt5}. 
Although in the former case it is difficult to speculate as to the origin of the instability, it is tempting to relate 
the same instability in the cascading plasma with the development of the chiral condensates responsible for the 
breaking of chiral symmetry. The fluctuations of such condensates are massive at high temperatures \cite{wip}. 
Exactly for this reason there is no high temperature regime for the deconfined cascading plasma with broken chiral symmetry 
--- correspondingly, 
there can not exist a black hole solution on the warped deformed conifold with fluxes \cite{ks} at high temperatures.
Of course, this does not exclude the possibility of such a black hole solution at low temperatures. 
The hydrodynamic stability of the symmetric phase all the way down to the deconfinement transition suggests though 
that the existence of the black hole in the broken phase would not modify the cosmological scenario 
proposed in \cite{bk}.
We return to these questions in more details in future work \cite{wip}.

\section*{Acknowledgments}
I would like to thank Banff International Research Station for hospitality 
where part of this work was done.
Research at Perimeter Institute is supported by the Government of
Canada through Industry Canada and by the Province of Ontario
through the Ministry of Research \& Innovation. I gratefully
acknowledge further support by an NSERC Discovery grant and support
through the Early Researcher Award program by the Province of
Ontario.

\end{document}